\documentclass[fleqn,usenatbib,useAMS]{mnras}

\usepackage{graphicx, color}% Include figure files

\usepackage{mathtools}
\usepackage{txfonts}

\usepackage[T1]{fontenc}
\usepackage{ae,aecompl}

\def\prl{Phys. Rev. Lett.}
\def\prd{Phys. Rev. D}

\def\apj{Astrophys. J.}

\def\mnras{Mon. Not. R. Astron. Soc.}

\def\Amax{\bar A^*_{\rm crit}}
\def\rsmax{\bar r_s^{\rm crit}}

\title[Dark matter heating of gas accreting onto Sgr A$^*$]{Dark matter heating of gas accreting onto Sgr A$^*$}
\author[Bennewitz, Gaidau, Baumgarte \& Shapiro]{Elizabeth R.~Bennewitz,$^{1}$ 
Cristian Gaidau,$^{2}$ 
Thomas W.~Baumgarte,$^{1}$ and \newauthor 
Stuart L.~Shapiro$^{2,3}$ \\
 $^{1}$Department of Physics and Astronomy, Bowdoin College, Brunswick, ME 04011\\
 $^{2}$Department of Physics, University of Illinois at Urbana-Champaign, Urbana, IL 61801\\
 $^{3}$Department of Astronomy and NCSA, University of Illinois at Urbana-Champaign, Urbana, IL 61801
}

\date{Accepted XXX. Received YYY; in original form ZZZ}

\pubyear{2019}

\begin{document}
\label{firstpage}
\pagerange{\pageref{firstpage}--\pageref{lastpage}}
\maketitle

%
%%%%%%%%%%%%%%%%%%%%%%%%%%%%%%%%%%%%%%%%%%%%%%%%%
\begin{abstract}
We study effects of heating by dark matter (DM) annihilation on black hole gas accretion.  We observe that, for
  reasonable assumptions about DM densities in spikes around supermassive 
  black holes, as well as DM masses
  and annihilation cross-sections within the standard WIMP model, heating by DM annihilation may have an
  appreciable effect on the accretion onto Sgr A$^*$ in the Galactic
  center.  Motivated by this observation we study the effects of such
  heating on Bondi accretion,
  i.e.~spherically symmetric, steady-state Newtonian accretion onto a black
  hole.  We consider different adiabatic indices for the gas, and
  different power-law exponents for the DM density profile.
  We find that typical transonic solutions with heating
  have a significantly reduced accretion rate.   However,
  for many plausible parameters, transonic solutions do not exist,
  suggesting a breakdown of the underlying assumptions of steady-state Bondi
  accretion.  
  Our findings indicate
  that heating by DM annihilation may play an important role
  in the accretion onto supermassive black holes at the center of
  galaxies, and may help explain the low accretion rate observed for
  Sgr A$^*$.
\end{abstract}
%%%%%%%%%%%%%%%%%%%%%%%%%%%%%%%%%%%%%%%%%%%%%%%%
%

%===============================================================================
\section{Introduction}
\label{sec:intro}
%===============================================================================

The spectacular images recently provided by the Event Horizon
Telescope (EHT) Collaboration \citep[see][as well as several follow-up
  publications]{EHT_I_19} have driven interest in accretion onto
supermassive black holes to new heights. One of the targets of the EHT
is Sgr A$^*$, the supermassive black hole with mass $M = 4 \times 10^6
M_\odot$ \citep{Ghezetal08,GenEG10,Giletal17} residing at the Galactic
center (GC). In this paper we are interested in the remarkably low
rate at which gas in the central bulge is actually accreting onto Sgr
A$^*$: it has long been recognized that this rate, estimated to be a
few times $\sim 10^{-8}M_{\odot}~{\rm yr}^{-1}$, is roughly three
orders of magnitude below the standard Bondi estimate for the rate at
which gas is gravitationally captured by the hole at $~0.1$ pc
\citep{Bagetal03,ShcB10,ResTQG17}.  The Bondi value for the rate is
determined from the gas density and temperature inferred from the
diffuse X-ray emission observed by {\it Chandra} at $\sim 2$ arc sec
(~$\sim 0.1$ pc) from the black hole and is $\sim 2 \times
10^{-5}M_{\odot}~{\rm yr}^{-1}$. The rate at which gas actually accretes onto
the black hole is inferred from polarization measurements
\citep{MarMZR07} and models of the near-horizon accretion flow and
emitted luminosity \citep{ShcB10,ResTQG17}.

The current explanation for this large difference begins with the
assumption that the gas originates from stellar winds from the $\sim
30$ Wolf-Rayet (WR) stars that orbit within $\sim 1$ pc from Sgr A$^*$
and that this gas thus has a broad distribution of angular momentum.
Hydrodynamic simulations in 3D \cite[see, e.g.][]{CuaNM08,ResQS18}
then suggest that, while the inflow rate at $\sim 0.1$ pc is $\sim 2-3
\times 10^{-5} M_{\odot}~{\rm yr}^{-1}$, which is close to the Bondi
value for the rate at which gas is gravitationally bound to the black
hole, only a small fraction of this mass actually accretes to smaller
radii $\ll 0.1$ pc, since only the low angular momentum tail of the
stellar wind is able to accrete. As it approaches the event horizon of
Sgr A$^*$, even this gas likely has sufficient angular momentum to
form a geometrically thick disk.  This near-horizon disk has been
simulated in general relativistic radiation-magnetohydrodynamics by
several investigators in recent years \citep[see, e.g.][and references
  therein]{SadWNAMC17,RyaRDTGQ17,ChaRNJS18}, forming the theoretical
framework for interpreting present and future observations of Sgr
A$^*$ by various instruments, including the EHT.

In this paper we investigate the possibility that heating by dark
matter (DM) annihilation may provide another reason why the accretion
rate onto Sgr A$^*$ is much lower than the canonical Bondi value. We
will explore this possibility by reconsidering the classic,
steady-state, spherical Bondi flow problem \citep{Bon52} but with
heating arising from the inclusion of DM annihilation \cite[see also][who 
found that the inclusion of thermal conduction in the Bondi solution similarly leads
to a reduction in the accretion rate]{JohQ07}.  Among the
parameters we allow to vary are the gas adiabatic index $\gamma$ and
the DM density profile parameter $\alpha$. The choice of $\gamma$
roughly accounts for cooling, which is not incorporated explicitly in
our equations: in the absence of heating, $\gamma = 5/3$ applies to
adiabatic flow (no cooling), while $\gamma =1$ applies to isothermal
flow (extreme cooling).  The parameter $\alpha$ is determined by the
power-law that describes the increase in the DM density with
decreasing radius $r$ from the GC.

Our goal is to use this simple, modified Bondi accretion model to
determine whether such heating can suppress the inflow rate for a
given set of gas dynamic parameters at large distance from the black
hole and a physically plausible DM annihilation rate. Even if
effective in reducing the accretion rate, it is not likely that
spherical Bondi flow will supplant our current understanding of the
more complicated flow patterns found in the 3D hydrodynamic
simulations.  However, if effective in the case of Bondi flow, heating
by DM annihilation may be another mechanism that should be
incorporated in future hydrodynamic simulations.  (In ``hot 
accretion" disk models like ADAFs, which also have been
employed to model Sag A*, heating by viscous
dissipation plays a dominant role; see, e.g., \cite{YuaN14} for a review.)

This paper is organized as follows. Section~\ref{sec:heating}
assembles plausible DM local and global parameters and uses them to
construct the heating rate due to DM
annihilation. Section~\ref{sec:basic_eqs} derives the basic Newtonian
equations for steady-state, spherical accretion of gas from rest at
infinity, incorporating this heating term.
Section~\ref{sec:transonic_results} identifies the range of parameters
for which the flow smoothly crosses a transonic point and summarizes
the accretion rates for such cases.
Section~\ref{sec:subsonic_results} does the same for solutions that
remain subsonic. Section~\ref{sec:Sgr_A} applies the results to the GC
and Sgr A$^*$.  We summarize our findings 
in Section~\ref{sec:summary}, and also
delineate some caveats that might alter the results obtained in the
earlier sections. 

%===============================================================================
\section{Heating rate due to DM annihilation}
\label{sec:heating}
%===============================================================================

We adopt the standard weakly
interacting massive particle (WIMP) model for the DM, which we treat
as collisionless particles of mass $m_{\chi}$ that undergo
annihilation reactions in a density spike around Sgr A$^*$.  The
heating rate per unit volume due to annihilation is given by
\begin{equation} \label{rate}
\Gamma(r) =\epsilon \frac{1}{2} n_{\rm DM}^2(r) \langle \sigma \varv\rangle 
              2 m_{\chi} c^2 
          =\epsilon \frac{\rho_{\rm DM}^2(r)}{m_{\chi}} \langle \sigma \varv\rangle c^2,
\end{equation}
where $n_{\rm DM}(r)$ is the DM number density, $\rho_{\rm DM}(r)$ is
the DM mass density, $\langle \sigma \varv\rangle$ is the annihilation
cross section, which we take to be constant (i.e. $s-$wave
annihilation), and $\epsilon$ is the efficiency at which the liberated
energy goes into the local heating of the accreting gas.   We take this 
   efficiency to be constant, even though in general
   it may also depend on the gas density and temperature,
   which enter the local opacity and optical depth
   to the annihilation product(s). Taking $\epsilon = 1$
   provides an upper limit to the heating rate and its influence on the
   flow.
If we follow
\cite{FieSS14} and adopt as our canonical DM annihilation cross
section and mass the reference point of \cite{Dayetal16} we then have
a DM particle with mass $m_{\chi}=35.25$~GeV annihilating to $b \bar
b$ with a cross section $\langle \sigma \varv \rangle = 1.7 \times
10^{-26} ~{\rm cm^2 s^{-1}}$, which are close to the values expected
for a thermal relic origin of DM. Appreciable $\sim 0.1-10$~GeV
gamma-ray emission is expected to accompany the annihilation process. For this
model, estimates of $\epsilon \sim {\cal O} (1)$ are not
unreasonable. We note that DM annihilation has been suggested as a
source of the $\sim 1-5$~GeV gamma-ray excess from the inner few
degrees of the GC observed by $\it Fermi$
\citep{Dayetal16,CalCW15,Fermi_16} and employed to assess the DM spike
and particle parameters \citep{FieSS14,SheSF15}, although other
plausible candidates for the excess (e.g.~a new population of pulsars)
have been proposed.

A supermassive black hole will steepen the density profile of DM
within the hole's sphere of influence, $r_s \approx GM/\varv_0^2$, which
is comparable to the region within which gas becomes bound to the
black hole.   We assume that the DM velocity dispersion $\varv_0$ in the GC
outside $r_s$ is comparable to the thermal velocity dispersion of the
gas. While the precise profile for this DM density spike depends on
the properties of DM and the formation history of the black hole, it
typically may be written as a piecewise power-law according to
\begin{equation} \label{spike}
\rho_{\it DM}(r) = \left\{
\begin{array}{ll} 
\rho_{\rm ann} (r_{\rm ann}/r)^{\rm \gamma_{\rm sp}},~~~~ & r \geq r_{\rm ann}, \\
\rho_{\rm ann} (r_{\rm ann}/r)^{\gamma_{\rm ann}},  & r < r_{\rm ann},
\end{array}
\right.
\end{equation}
%\begin{eqnarray}
%\label{spike}
%\rho_{\it DM}(r) &=& \rho_{\rm ann} (r_{\rm ann}/r)^{\rm \gamma_{\rm sp}}, 
%                  \ \ \ r \geq r_{\rm ann}, \nonumber \\
%              &=& \rho_{\rm ann} (r_{\rm ann}/r)^{\gamma_{\rm ann}}, \ \ \ r < r_{\rm ann}. 
%\end{eqnarray}
plunging to near zero in the vicinity of the black hole horizon.
If, for example, the supermassive black hole grows adiabatically from
a smaller seed \citep{Pee72}, before which the DM density obeyed a
generalized Navarro-Frenk-White profile \citep[NFW,][]{NavFW97} of the
form $\rho_{\rm DM} \sim r^{-\gamma_{\rm c}}$, then the black hole
will modify the profile, forming a spike given by eqn.~(\ref{spike})
with $\gamma_{\rm sp} = (9-2\gamma_{\rm c})/(4-\gamma_{\rm c})$
\citep{GonS99}.  Possible values for $\gamma_{\rm c}$ and $\gamma_{\rm
  sp}$ are reviewed in \cite{FieSS14} and references therein, but here
we choose as a canonical value $\gamma_{\rm c} =1$, for which
$\gamma_{\rm sp} = 7/3$.  We note that for $0 < \gamma_{\rm c} \leq 2$
the power law $\gamma_{\rm sp}$ varies at most between 2.25 and 2.50
for this adiabatic growth scenario.  By contrast, gravitational
scattering off a dense stellar component inside $r_s$ could heat
the DM, softening the spike profile and ultimately driving it to a
final equilibrium value of $\gamma_{\rm sp} = 3/2$
\citep{Mer04,GneP04} or even to disruption \citep{WanBVW15}; we will
therefore show results for a range of different values of
$\gamma_{\rm sp}$.

% also show some results for $\gamma_{\rm sp} = 3/2$ as well as $7/3$ below.

At $r=r_{\rm ann}$ the DM density in the spike reaches $\rho_{\rm
  ann}$, once referred to as the ``annihilation plateau" density. At
this radius the annihilation time scale equals the Galaxy age $T$,
whereby
\begin{equation}
\label{annihilation}
\rho_{\rm ann} = \frac{m_{\chi}}{\langle \sigma \varv\rangle T}.
\end{equation} 
For $r < r_{\rm ann}$ the density in the spike is not a flat plateau
profile but varies as in eqn.~(\ref{spike}) with $\gamma_{\rm ann} =
1/2$ for $s$-wave annihilation \citep{Vas07,ShaS16}.  For our
canonical particle model and $T \approx 10^{10}$~yr, we find
$\rho_{\rm ann} = 1.7 \times 10^8 M_{\odot}~{\rm pc^{-3}}$ and $r_{\rm
  ann} = 3.1 \times 10^{-3}$~pc.

{\it Chandra} X-ray measurements at approximately
$2''$ from the GC give thermal temperatures $kT \approx 1.3$~keV,
corresponding to sound speeds $a_s = (\gamma kT/\mu m_p)^{1/2} \approx
550$~km/s, assuming $\gamma = 5/3$ and a mean molecular weight $\mu =
0.7$ \citep{Bagetal03}.  For a black-hole mass of $M \sim 4 \times
10^{6} M_\odot$ this yields a Bondi capture radius $R_B = GM/a_s^2
\sim 0.061$~pc $\sim r_s$.

For radii $r \geq r_{\rm ann}$ we may write the heating rate in eqn.~(\ref{rate}) 
as a power-law,
\begin{equation} \label{rate2}
\Gamma(r) = \Gamma_0 \left( \frac{r_{\rm ann}}{r} \right)^{2 \gamma_{\rm sp}}, 
\ \ \ \Gamma_0= \epsilon \frac{\rho_{\rm ann}^2}{m_{\chi}} \langle \sigma \varv\rangle c^2,
\ \ \ r \geq r_{\rm ann}.
\end{equation}
For our canonical DM model we find
$\Gamma_0 = \epsilon \times 3.35 \times 10^{-11} \mbox{ergs}\,
\mbox{cm}^{-3}\, \mbox{s}^{-1}$.  In our discussion of heated Bondi
accretion in the following sections we will ignore the transition from
$\gamma_{\rm sp}$ to $\gamma_{\rm ann}$ at $r = r_{\rm ann}$, and
will, for simplicity, assume that the heating is governed by
(\ref{rate2}) at all radii.  Typically the gas accretion rate is 
established near $r_s \gg r_{\rm ann}$, justifying our simplification.
While it is straight-forward to relax
this assumption, it leads to a well-defined mathematical problem with
few free parameters; we will comment when this assumption may affect
our astrophysical conclusions.

We recall that for typical values of $1 \lesssim \gamma \lesssim 5/3$
and $\gamma_{\rm sp}$ the rate at which mass is captured by the black
hole in smooth, transonic Bondi flow in the absence of heating is
established by gas parameters near the transonic point $r_s \sim R_B
\gg r_{\rm ann}$. The steady-state rate of capture and spherical
accretion in this case is given by 
  ${\dot M}_0 = 4 \pi r^2 \rho u \sim 4
\pi  \lambda_s \rho_s (GM)^2/a_s^3$, which is independent of $r$.  Here
$\lambda_s$ is a parameter of order unity depending on $\gamma$ (see
Eq.~(\ref{lambda_s}) below).  The
corresponding gas density inside $r_s$ increases as $\rho(r) \sim
\rho_s (r_s/r)^{3/2}$ and the square of the sound speed increases as
        $a^2(r) \sim a_s^2 (r_s/r)^{3(\gamma-1)/2}$.  The importance of heating by DM annihilations
    may then be inferred from the ratio $\cal R$ of the heating rate
    by DM annihilation in a volume between radius $r/2$ and $r$ over
    the rate at which thermal energy in an unheated gas would flow
    adiabatically into this volume:
%\begin{eqnarray}
\begin{align} \label{ratio}
{\cal R}(r) &\sim \frac{ \Gamma(r) 4 \pi r^3}{{\dot M_0} a^2(r)}  \nonumber \\
&\sim \frac{4 \pi \Gamma_0 r_s^3}{\dot M_0} 
\left( \frac{r_{\rm ann}}{r_s} \right)^{2 \gamma_{\rm sp}}
\left( \frac{r_s}{GM} \right)
\left( \frac{r}{r_s} \right)^{3(\gamma+1)/2-2\gamma_{\rm sp}} 
\end{align}
%\end{eqnarray}
Evaluating this ratio at $r_s$ for our canonical DM model with ${\dot
  M}_0 \sim 10^{-5}M_{\odot}~{\rm yr}^{-1}$, $\epsilon \sim 1$ and
$\gamma_{\rm sp} = 7/3$ gives ${\cal R}(r_s) \sim 1$, i.e.~${\cal R}(r_s)$ is of order
unity.  Note also that this ratio increases with decreasing $r$
whenever $\gamma < 4 \gamma_{\rm sp}/3 - 1$, which is the case for all realistic
values of $\gamma$ when $\gamma_{\rm sp} = 7/3$ (but not when
$\gamma_{\rm sp} = 3/2$).  The fact that ${\cal R}$ is of order unity
at the sonic radius and may grow to even larger values at smaller
radii suggests that DM annihilation heating, if present, will
significantly affect the inflow solution.
This observation motivates our study of the effects of this heating on
the simplest possible accretion model, namely spherical Bondi accretion.

%===============================================================================
\section{Basic equations}
\label{sec:basic_eqs}
%===============================================================================

%===============================================================================
\subsection{Fluid equations}
\label{sec:fluid_eqs}
%===============================================================================

Bondi accretion \cite{Bon52} describes the spherically symmetric
steady-state accretion of a fluid onto a black hole, from rest at infinity.  Following
Bondi's original work we will adopt a Newtonian treatment here (see
\cite{Mic72a,Sha73a} or \cite{ShaT83}, hereafter ST, for relativistic
generalizations), and will describe the black hole as a point-mass
$M$, generating a Newtonian potential $GM/r$, where $r$ is
the distance from the black hole.  The fluid flow is then governed by
the Newtonian fluid equations -- the first law of thermodynamics, the
continuity equation, and the Euler equation -- in the presence of this
potential.  Unlike Bondi, however, we will not assume that the fluid
flow is adiabatic, and will instead allow for a heating term $\Gamma$,
as discussed in Sec.~\ref{sec:heating}.

%===============================================================================
\subsubsection{Equations in differential form}
\label{sec:diff_eqs}
%===============================================================================

In the presence of a heating term $\Gamma$, the first law of
thermodynamics takes the form
\begin{equation} \label{first_law}
\frac{d \epsilon}{dt} + P \frac{d}{dt} \left( \frac{1}{\rho} \right) 
= \frac{\Gamma}{\rho}, 
\end{equation}
where $\epsilon$ is the specific internal energy density, $\rho$ the
mass density, and $P$ the pressure.  The time derivatives in
Eq.~(\ref{first_law}) are to be taken along the fluid flow, e.g.
\begin{equation} \label{time_deriv}
\frac{d\epsilon}{dt} = \frac{\partial \epsilon}{\partial t} 
+ \varv^r \frac{\partial \epsilon}{\partial r},
\end{equation}
where we have assumed spherical flow, and where $\varv^r$ is the radial
component of the fluid velocity.  We assume that, as $r
\rightarrow \infty$, the fluid is at rest, $\varv^r \rightarrow 0$, at uniform 
density $\rho \rightarrow \rho_\infty$.

We will adopt a gamma-law equation of state (EOS) throughout, so that 
\begin{equation} \label{eos}
P = (\gamma - 1) \rho \epsilon.
\end{equation}
For adiabatic flow, the constant $\gamma$ can be
related to the specific heat of the gas.  For a nonrelativistic, ideal 
monatomic gas, which
is relevant for the accretion problems we study here, we have
$\gamma = 5/3$.  Even for the nonadiabatic flows considered here we
always assume that $\gamma$ remains constant throughout; we will pay
special attention to $\gamma = 5/3$, but will consider other values
also to account for cooling.  We define $K \equiv P \rho^{-\gamma}$, so that
\begin{equation} \label{polytropic}
P = K \rho^\gamma.
\end{equation}
In the adiabatic case, i.e.~for isentropic flow, $K$ is a constant
(see Eq.~(\ref{K_prime_Gamma}) below), but in general that is not the
case.  We can then compute the sound speed $a$ from
\begin{equation} \label{a}
a^2 = \left. \frac{dP}{d\rho} \right|_s= \gamma K \rho^{\gamma - 1} = \gamma \frac{P}{\rho},
\end{equation}
where the derivative in the second term is taken at constant entropy
$s$, and hence at constant $K$.

For spherically symmetric flow, the continuity equation can be written as
\begin{equation} \label{continuity_1}
\frac{\partial \rho}{\partial t} + 
\frac{1}{r^2} \frac{\partial}{\partial r} (r^2 \rho \varv^r) = 0,
\end{equation}
while the Euler equation becomes
\begin{equation} \label{euler_1}
\frac{\partial \varv^r}{\partial t} + 
\varv^r \frac{\partial \varv^r}{\partial r} = 
-  \frac{1}{\rho} \frac{\partial P}{\partial r}  - \frac{GM}{r^2},
\end{equation}
where we have assumed that the fluid's self-gravity can be ignored.

%===============================================================================
\subsubsection{Equations for steady-state flow}
\label{sec:steady_state}
%=============================================================================q==

We now focus on steady state, so that all partial derivatives with
time vanish.  Since we will mostly be concerned with in-flow, we also
define
\begin{equation}
u = - \varv^r
\end{equation}
for convenience.  The first law (\ref{first_law}) can then be written as
\begin{equation} \label{first_law_2}
\frac{d \epsilon}{dr} + P \frac{d}{dr} \left( \frac{1}{\rho} \right) 
= - \frac{\Gamma}{\rho u}.
\end{equation}
Combining this with (\ref{eos}) and (\ref{polytropic}) we find
\begin{equation} \label{K_prime_Gamma}
\frac{dK}{dr} = - \frac{(\gamma - 1)}{\rho^\gamma} \frac{\Gamma}{u}.
\end{equation}
As expected, $K$ becomes a constant for adiabatic flow, when $\Gamma = 0$.

For steady-state flow, the continuity equation (\ref{continuity_1}) reduces to
\begin{equation} \label{continuity_2a}
\frac{d}{d r} (r^2 \rho u) = 0,
\end{equation}
or, equivalently,
\begin{equation} \label{continuity_2}
\frac{\rho'}{\rho} +  \frac{u'}{u} + \frac{2}{r}  = 0,
\end{equation}
where a prime denotes a derivative with respect to $r$.  Finally, the
Euler equation (\ref{euler_1}) becomes
\begin{equation}  \label{euler_2}
u u' = - \frac{1}{\rho} P' - \frac{GM}{r^2}.
\end{equation}
In order to eliminate the pressure $P$ from this equation we take a
derivative of (\ref{polytropic}),
\begin{equation}
P' = \left. \frac{\partial P}{\partial \rho} \right|_K \rho' + 
\left. \frac{\partial P}{\partial K} \right|_\rho K'
 = a^2 \rho' + \rho^\gamma K',
\end{equation}
and insert this into (\ref{euler_2}) to obtain
\begin{equation} \label{euler_3}
u u' = - a^2 \frac{\rho'}{\rho} - \rho^{\gamma -1} K' - \frac{GM}{r^2}.
\end{equation}
Using (\ref{K_prime_Gamma}) we can now eliminate $K$ and find
\begin{equation} \label{euler_3a}
u u' = - a^2 \frac{\rho'}{\rho} + (\gamma - 1) \frac{\Gamma}{\rho u}  
- \frac{GM}{r^2}.
\end{equation}

Eqs.~(\ref{K_prime_Gamma}), (\ref{continuity_2}) and (\ref{euler_3a})
form a coupled system of three ordinary differential equations for the
dependent variables $K$, $\rho$ and $u$ describing the nonadiabatic
fluid flow profiles (note that $K$ couples to $u$ and $\rho$ 
through Eq.~(\ref{a})).  The last two of these equations contain both
$u'$ and $\rho'$; it is therefore convenient to combine the equations
and find expressions for $u'$ and $\rho'$ alone.  This results in
\begin{equation} \label{u_prime}
u' = u \frac{D_1 + H}{D}
\end{equation}
and
\begin{equation} \label{rho_prime}
\rho' = - \rho \frac{D_2 + H}{D},
\end{equation}
where we have defined the coefficients
\begin{align} % \label{coefficients}
D_1 & \equiv \displaystyle \frac{2 a^2}{r} -\frac{GM}{r^2} \label{D_1} \\
D_2 & \equiv \displaystyle \frac{2 u^2}{r} -\frac{GM}{r^2} \label{D_2} \\
D & \equiv u^2 - a^2 \label{D}\\
H & \equiv % (\gamma - 1)  A^* \left( \frac{r_{\rm ann}}{r} \right)^\alpha
(\gamma - 1)  \frac{\Gamma}{\rho u} \label{H} 
\end{align}

%===============================================================================
\subsubsection{Integrated equations}
\label{sec:int_eqs}
%===============================================================================
      
Both the continuity equation and the Euler equation can also be
integrated directly.  Integrating the continuity equation
(\ref{continuity_2a}) yields
\begin{equation} \label{M_dot}
\dot M = 4 \pi \rho u r^2 = \mbox{~constant~},
\end{equation}
where $\dot M$ is the accretion rate.  

Integrating the first term on the right-hand side of the Euler
equation (\ref{euler_3a}) yields
\begin{align} \label{int_first_term}
% \begin{array}{rcl}
\displaystyle \int a^2 \frac{\rho'}{\rho} dr & 
=  \displaystyle \gamma \int \frac{P}{\rho^2} d\rho 
= \gamma \int K \rho^{\gamma - 2} d\rho 
=  \displaystyle \frac{\gamma}{\gamma - 1} \int K d \rho^{\gamma - 1} 
\nonumber \\[3mm]
& =  \displaystyle \frac{\gamma}{\gamma - 1} \left[ K \rho^{\gamma - 1} \right] 
- \frac{\gamma}{\gamma - 1} \int \rho^{\gamma - 1} dK \nonumber \\[3mm]
& =  \displaystyle \left[ \frac{a^2}{\gamma - 1} \right] 
+ \gamma \int \frac{\Gamma}{\rho u} dr,
% \end{array}
\end{align}
where we have used (\ref{a}), (\ref{polytropic}), integration by
parts, and (\ref{K_prime_Gamma}).  Integrating the remaining terms in
(\ref{euler_3a}) and using (\ref{int_first_term}) we now obtain
\begin{equation} \label{Bernoulli_almost}
\frac{u^2}{2} + \frac{a^2}{\gamma - 1} - \frac{GM}{r} 
+ \int_\infty^r \frac{\Gamma}{\rho u} dr = \frac{a_{\infty}^2}{\gamma - 1}
\end{equation}
where $a_\infty$ is the sound speed at $r \rightarrow \infty$.  In
order to integrate the heating term we now write $\Gamma$ as
\begin{equation} \label{A_def}
\Gamma = \rho u A^* \left( \frac{r_{\rm ann}}{r} \right)^{\alpha}
\end{equation}
where $A^*$ becomes a constant if $\alpha$ is chosen as in
(\ref{alpha}) below.  To see this, we combine (\ref{A_def}) with
(\ref{rate2}) and solve for $A^*$,
\begin{equation}
% \begin{array}{rcl}
A^* = \displaystyle \frac{\Gamma_0}{\rho u} \left( \frac{r_{\rm ann}}{r} \right)^{2 \gamma_{\rm sp} - \alpha} 
% = \displaystyle \frac{\Gamma_0}{\rho u} \left( \frac{r_{\rm ann}}{r} \right)^{2 \gamma_{\rm sp} - \alpha} \\[3mm]
= \displaystyle \frac{4 \pi r_{\rm ann}^2 \Gamma_0}{\dot M} \left( \frac{r_{\rm ann}}{r} \right)^{2 \gamma_{\rm sp} - \alpha - 2},
% \end{array}
\end{equation}
where we have used (\ref{M_dot}) in the last step.  We now choose
\begin{equation} \label{alpha}
\alpha \equiv 2 \gamma_{\rm sp} - 2
\end{equation}
so that $A^*$ becomes the constant
\begin{equation} \label{A_star}
A^* = \displaystyle \frac{4 \pi r_{\rm ann}^2 \Gamma_0}{\dot M}.
\end{equation}
Since $\Gamma_0$ has units of energy per time and volume, $A^*$ has
units of length per time squared, or, equivalently, speed squared per
length.  For $\gamma_{\rm sp} = 7/3$ we now have $\alpha = 8/3$, and
for $\gamma_{\rm sp} = 3/2$ we find $\alpha = 1$.  Since $A^*$ depends
on the accretion rate $\dot M$, it cannot be computed from the DM model 
parameters of Section \ref{sec:heating} alone.  We will use representative values 
in many of our examples, and will evaluate possible values of $A^*$ for Sgr $A^*$
in Section \ref{sec:Sgr_A} below.

Inserting (\ref{A_def}) into (\ref{Bernoulli_almost}), and assuming
$\alpha > 1$, we can now integrate the heating term and obtain the
Bernoulli equation
\begin{equation} \label{Bernoulli}
\frac{u^2}{2} + \frac{a^2}{\gamma - 1} - \frac{GM}{r} 
- \frac{A^*}{\alpha - 1} \frac{r_{\rm ann}^\alpha}{r^{\alpha - 1}} 
= \frac{a_{\infty}^2}{\gamma - 1}.
\end{equation}
For $\alpha = 1$ the integral of the heating term diverges
logarithmically as $r \rightarrow \infty$.  For astrophysical models
of galactic centers this may not be a problem, since the accretion
flow does not extend to arbitrarily large distances.  For our
treatment here, however, we will take $\alpha > 1$.  %, but will
% consider values of $\alpha$ close to unity.

Inserting (\ref{A_def}) into (\ref{K_prime_Gamma}) yields
\begin{equation} \label{K_prime}
K' = - (\gamma -1) \frac{A^*}{\rho^{\gamma - 1}}  \left( \frac{r_{\rm ann}}{r} \right)^\alpha,
\end{equation}
which cannot be integrated analytically unless $A^* = 0$.

%===============================================================================
\subsection{Adiabatic flow  revisited:  $\Gamma = 0 = A^*$}
\label{sec:adiabatic_flow}
%===============================================================================

Before embarking on a treatment of heated Bondi accretion in the
following sections, we first review the special case of adiabatic flow
with $\Gamma = 0 = A^*$.  We refer to ST for a review and derivation, and 
summarize only the most important results here.

We start by distinguishing between {\em subsonic} and {\em transonic}
solutions.  Subsonic solutions, for which $u < a$ everywhere, can have
arbitrary accretion rates $\dot M$ up to a certain maximum value
$\dot M_0$, which will be given by the transonic accretion rate
discussed below.  We can express this accretion rate as
\begin{equation} \label{M_dot_2}
\dot M = 4 \pi \lambda \rho_\infty a_\infty \left( \frac{GM}{a_\infty^2} \right)^2
\end{equation}
with $\lambda < \lambda_s$, where the maximum value $\lambda_s$ is
given by (\ref{lambda_s}) below.  For a given accretion rate,
Eqs.~(\ref{M_dot}) and (\ref{Bernoulli}) together with (\ref{a}) then
provide three equations for the three unknowns $u$, $\rho$ and $a$ as a
function of radius $r$.  Solving these three equations provides
algebraic equations that describe the fluid profiles everywhere.

For transonic solutions we must have $u = a$ at some {\em sonic
  radius} $r_s$, implying that the coefficient $D$ vanishes at this
point (see Eq.~(\ref{D})).  This, in turn, implies that $D_1$ and
$D_2$, which become identical when $u = a$, also have to vanish at
$r_s$, since otherwise the solutions $u$ and $\rho$ to (\ref{u_prime})
and (\ref{rho_prime}) cannot be regular.  The conditions $D_1 = 0$ and
$u = a$ together with Eqs.~(\ref{a}), (\ref{M_dot}), (\ref{Bernoulli})
evaluated at $r = r_s$ provide five equations that can now be solved
for $r_s$, $u_s$, $\rho_s$, $a_s$ and $\dot M$.  Requiring regularity
determines the sonic radius, given by
\begin{equation} \label{r_s_adiabatic}
r_s = \left( \frac{5 - 3 \gamma}{4} \right) \frac{GM}{a_\infty^2}
\end{equation}
(see Eq.~(14.3.14) in ST) and yields a unique accretion rate
$\dot M_0$, given by (\ref{M_dot_2}) with
\begin{equation} \label{lambda_s}
\lambda = \lambda_s = \left( \frac{1}{2} \right)^{(\gamma + 1)/2(\gamma -1)} 
\left( \frac{5 - 3 \gamma}{4} \right)^{-(5-3\gamma)/2(\gamma - 1)}
\end{equation}
and $\lambda_s = 1/4$ in the limit of $\gamma = 5/3$ (see
Eq.~(14.3.17) in ST).  As we discussed above, this accretion rate
$\dot M_0$ also determines the maximum possible accretion rate for
subsonic flows.

While a Newtonian treatment allows both subsonic and supersonic flows,
i.e.~all accretion rates (\ref{M_dot_2}) with $\lambda \leq
\lambda_s$, a relativistic treatment allows only the transonic
solution with $\lambda = \lambda_s$ for regularity everywhere outside 
the black hole (see Appendix G in ST).  Since we
expect that a similar treatment carries over to heated Bondi
accretion, we will be primarily interested in transonic solutions
whenever they exist for smooth steady-state flow.  We also note that,
for $\gamma = 5/3$, Eq.~(\ref{r_s}) indicates that the sonic radius
vanishes, $r_s = 0$.  This is an artifact of our Newtonian treatment;
in a relativistic treatment the sonic radius for $\gamma = 5/3$ is
instead given by
\begin{equation}
r_s \approx \frac{3}{4} \frac{GM}{a_\infty^2}
\end{equation} 
(see Exercise G.1 in ST).  For nonrelativistic thermal speeds at
large distances, $r_s \gg GM/c^2$, so that
relativistic corrections to the Newtonian accretion rate are small.

%===============================================================================
\subsection{Nondimensional equations} 
\label{sec:nondimensional}
%===============================================================================

Before proceeding it is useful to cast the key equations in
nondimensional form.  To do so, we express the fluid variables in
terms of asymptotic values
\begin{equation} \label{fluid_bar}
a = a_\infty \bar a,~~~~~~~~  u = a_\infty \bar u,~~~~~~~~~~\rho = \rho_\infty \bar \rho,
\end{equation}
where the ``barred" variables are now dimensionless.  The radius
\begin{equation} \label{r_a}
r_a \equiv \frac{GM}{a_\infty^2}
\end{equation} 
then defines a natural length-scale, motivating the rescaling
\begin{equation}
r = r_a \bar r.
\end{equation}
In particular we have $\bar r_{\rm ann} = r_{\rm ann} / r_a \simeq
0.0507$, where we adopted $r_{\rm ann}
\simeq 3.1 \times 10^{-3}$ pc and $r_a \simeq r_s \simeq 0.061$ pc
       as discussed in Section
       \ref{sec:heating}.

We similarly write
\begin{equation} \label{bar_P_K}
P = P_\infty \bar P,~~~~~~~~~ K = K_\infty \bar K
\end{equation}
and identify from (\ref{a})
\begin{equation}
P_\infty = \frac{a_\infty^2 \rho_\infty}{\gamma},~~~~~~~~~~~~ K_\infty = \frac{a_\infty^2}{\gamma \rho_\infty^{\gamma - 1}}.
\end{equation}
In terms of these quantities Eq.~(\ref{a}) yields
\begin{equation} \label{a_nd}
\bar a^2 = \bar K \bar \rho^{\gamma - 1} = \frac{\bar P}{\bar \rho}.
\end{equation}
Finally we rescale $A^*$ according to
\begin{equation} \label{A_star_nd}
A^* = \frac{a_\infty^2}{r_a} \bar A^*.
\end{equation}

In terms of our nondimensional variables, Eqs.~(\ref{u_prime}) and
(\ref{rho_prime}) become
\begin{equation} \label{u_prime_nd}
\bar u' = \bar u \frac{\bar D_1 + \bar H}{\bar D}
\end{equation}
and
\begin{equation} \label{rho_prime_nd}
\bar \rho' = - \bar \rho \frac{\bar D_2 + \bar H}{\bar D},
\end{equation}
where the primes now denote a derivative with respect to $\bar r$, and
where the coefficients are now given by
\begin{align} % \label{coefficients_nd}
\bar D_1 & \equiv  \displaystyle \frac{2 \bar a^2}{\bar r} -\frac{1}{\bar r^2} \label{D_1_nd} \\
\bar D_2 & \equiv  \displaystyle \frac{2 \bar u^2}{\bar r} -\frac{1}{\bar r^2} \label{D_2_nd} \\
\bar D & \equiv  \bar u^2 - \bar a^2 \label{D_nd}\\
\bar H & \displaystyle \equiv  (\gamma - 1)  \bar A^* \left( \frac{\bar r_{\rm ann}}{\bar r} \right)^\alpha
\end{align}
Eq.~(\ref{K_prime}) becomes
\begin{equation} \label{K_prime_nd}
\bar K' = - \gamma (\gamma -1) \frac{\bar A^*}{\bar \rho^{\gamma - 1}}  \left( \frac{\bar r_{\rm ann}}{\bar r} \right)^\alpha,
\end{equation}
where we note the appearance of an extra factor of $\gamma$, which
arises due to the definition of $\bar K$ in (\ref{bar_P_K}).

We also write the integrated continuity equation (\ref{M_dot}) as 
\begin{equation}
\dot M = 4 \pi \bar \rho \bar u \bar r^2 \rho_\infty a_\infty 
\left( \frac{G M}{a_\infty^2} \right)^2 
= \frac{\bar \rho \bar u \bar r^2}{\lambda_s} \dot M_0
= \dot{\bar M} \dot M_0,
\end{equation}
where we have used (\ref{M_dot_2}) with $\lambda = \lambda_s$ 
for $\dot M_0$, and where we identify
\begin{equation} \label{M_dot_nd}
\dot {\bar M} = \frac{\bar \rho \bar u \bar r^2}{\lambda_s}.
\end{equation}
Finally, the Bernoulli equation (\ref{Bernoulli}) now takes the form
\begin{equation} \label{Bernoulli_nd}
\frac{\bar u^2}{2} + \frac{\bar a^2}{\gamma - 1} - \frac{1}{\bar r} - \frac{\bar A^*}{\alpha - 1} \frac{\bar r_{\rm ann}^\alpha}{\bar r^{\alpha - 1}} = \frac{1}{\gamma - 1}.
\end{equation}

%===============================================================================
\section{Heated transonic flow}
\label{sec:transonic_results}
%===============================================================================

%===============================================================================
\subsection{Computational strategy}
\label{sec:transonic_strategy}
%===============================================================================

Before discussing results for heated transonic flow we first outline our
computational strategy.

For transonic flow there exists (at least) one sonic radius
$\bar r_s$ at which $\bar u = \bar a$.  In the following we will
denote physical quantities evaluated at this radius with a subscript
$s$, e.g.~$\bar u_s = \bar a_s$.  At $\bar r_s$, the denominator
$\bar D$ in Eqs.~(\ref{u_prime_nd}) and (\ref{rho_prime_nd}) vanishes,
so that, for regular solutions to exist, the numerators have to vanish 
as well.  This implies
\begin{equation} \label{a_s}
\bar a_s^2 = \frac{1}{2 \bar r_s} - \frac{\gamma - 1}{2} \bar A^* \frac{\bar r_{\rm ann}^\alpha}{\bar r_s^{\alpha - 1}}.
\end{equation} 
Inserting this expression into the Bernoulli equation
(\ref{Bernoulli_nd}), evaluated at $\bar r = \bar r_s$, yields
\begin{equation} \label{r_s}
f(\bar r_s) \equiv \frac{5 - 3 \gamma}{4} \bar r_s^{\alpha - 2} - \beta \bar A^* \bar r_{\rm ann}^\alpha - \bar r_s^{\alpha - 1} = 0
\end{equation}
where we have abbreviated
% [CHECK: we may need to introduce $\beta_1$
% and $\beta_2$ separately, as in Elizabeth's thesis]
\begin{equation} \label{beta}
\beta \equiv (\gamma - 1) \left( \frac{1}{4} (\gamma + 1) + \frac{1}{\alpha - 1} \right).
\end{equation}
We note that $\beta > 0$ for all values of $\gamma > 1$ and $\alpha >
1$.  Eq.~(\ref{r_s}) now determines the sonic radius $\bar r_s$; in
the adiabatic limit $\bar A^* = 0$ we recover (\ref{r_s_adiabatic}) in
nondimensional form.  In general, when $\alpha$ is not an integer, we
have to solve Eq.~(\ref{r_s}) numerically with a root-finding method.
Given $\bar r_s$, we can then find $\bar a_s = \bar u_s$ from
(\ref{a_s}).

Since, in the presence of heating, we cannot integrate
(\ref{K_prime_nd}) analytically, we cannot obtain a closed-form
expression for $\bar K_s$.  We instead employ an iterative ``shooting"
method, by which we guess a value of $\bar K_s$, and then integrate
(\ref{K_prime_nd}) together with (\ref{u_prime_nd}) and
(\ref{rho_prime_nd}) from $\bar r = \bar r_s$ to some large value
$\bar r_{\rm out} \gg \bar r_a$.  At $\bar r_{\rm out}$ we compare the
integrated values of $\bar K$, $\bar u$ and $\bar \rho$ with the
boundary conditions $\bar u_\infty = 0$ and $\bar K_\infty = \bar
\rho_\infty = 1$, and adjust $\bar K_s$ to obtain better agreement.

We employ l'H\^opital's rule to evaluate eqs.~(\ref{u_prime_nd})
and (\ref{rho_prime_nd}) directly at $\bar r_s$.
Specifically, we take
derivatives with respect to $\bar r$ of both the numerator and
denominator of Eq.~(\ref{u_prime_nd}), using (\ref{a_nd}) to express
derivatives of $\bar a$ in terms of $\bar \rho$, and the continuity
equation to express the latter in terms derivatives of $\bar u$.  The
result is a quadratic equation for $\bar u'$.  When this equation has 
two real solutions, one solution describes inflow whereas the other
solution describes outflow (wind) solutions.  We pick the former, in 
practice choosing that solution for which $\bar u'$ is smaller than 
$\bar a'$, so that our solutions are subsonic outside $\bar r_s$.

Once $\bar K_s$ has been found, we can also find $\bar \rho_s$ from
(\ref{a_nd}), and then the accretion rate $\dot {\bar M}$ from
(\ref{M_dot_nd}), evaluated at $\bar r_s$.  Finally,
Eqs.~(\ref{K_prime_nd}) together with (\ref{u_prime_nd}) and
(\ref{rho_prime_nd}) can also be integrated inwards, thereby providing
fluid flow profiles inside the sonic radius.

In the following Sections we will discuss the individual steps in this
procedure for specific choices of the parameters $\alpha$ and
$\gamma$.

%===============================================================================
\subsection{Finding the sonic radius}
\label{sec:find_r_s}
%===============================================================================

\begin{figure*}
\begin{center}
\includegraphics[width=0.3\textwidth]{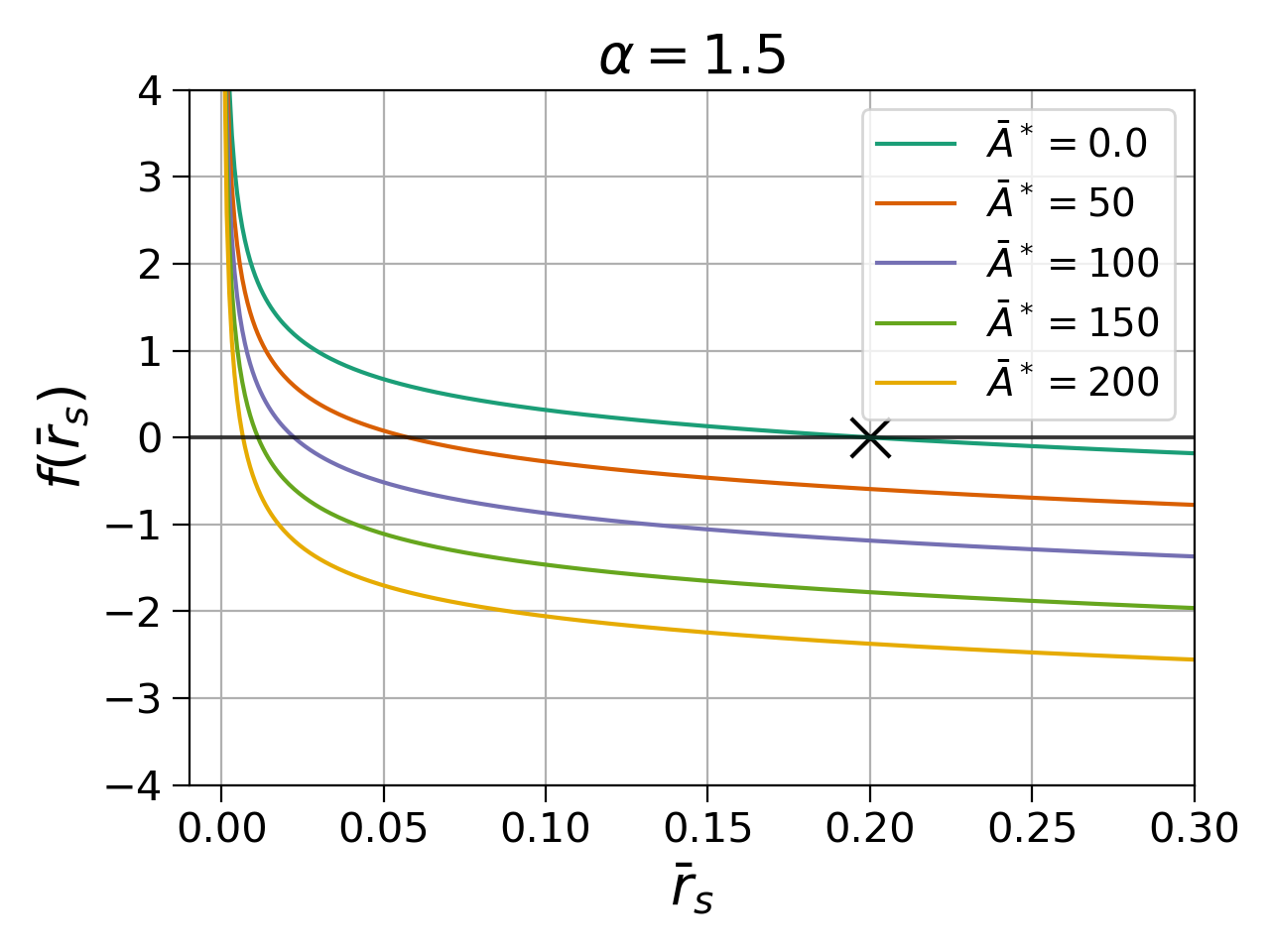}
\includegraphics[width=0.3\textwidth]{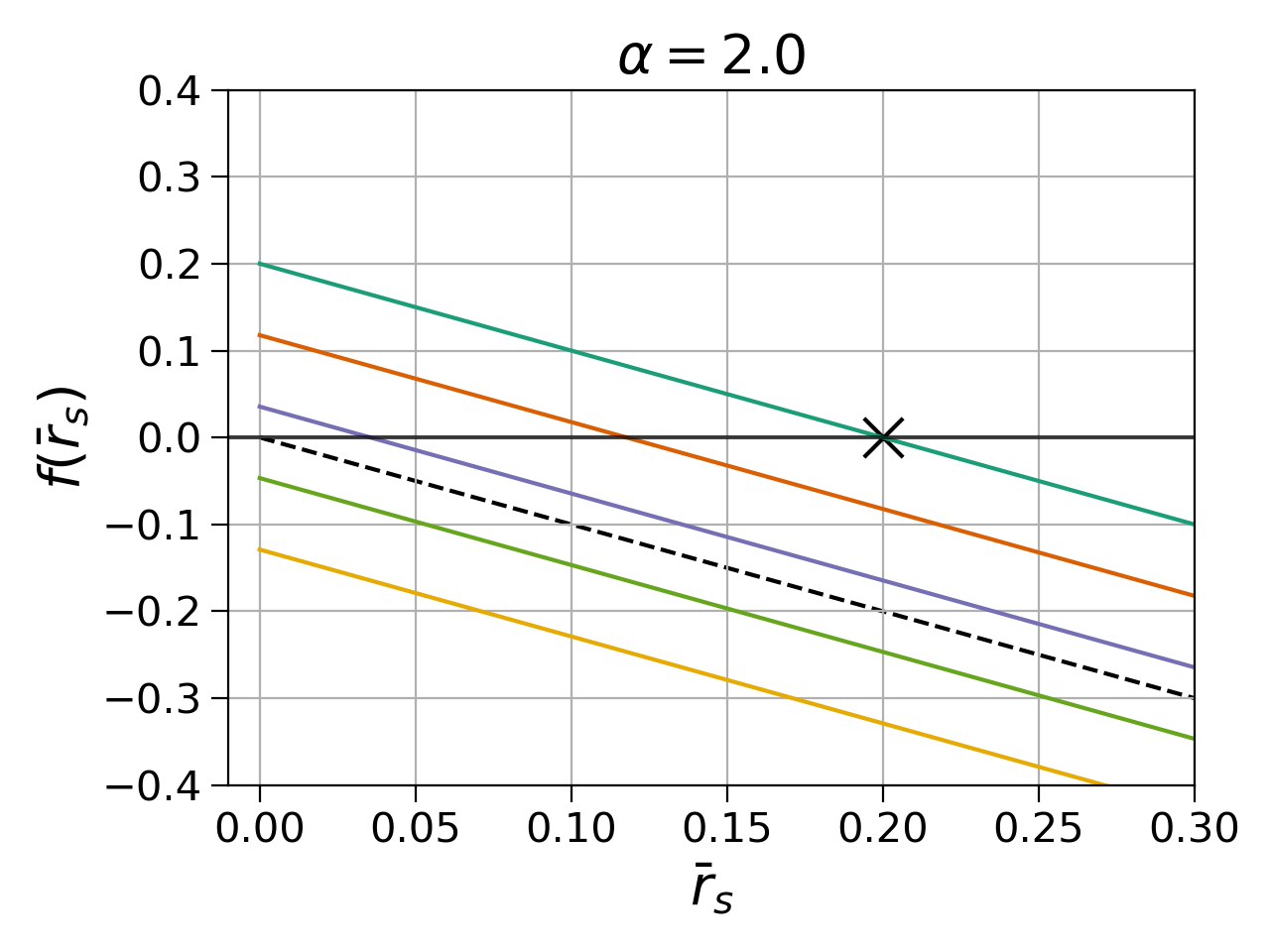}
\includegraphics[width=0.3\textwidth]{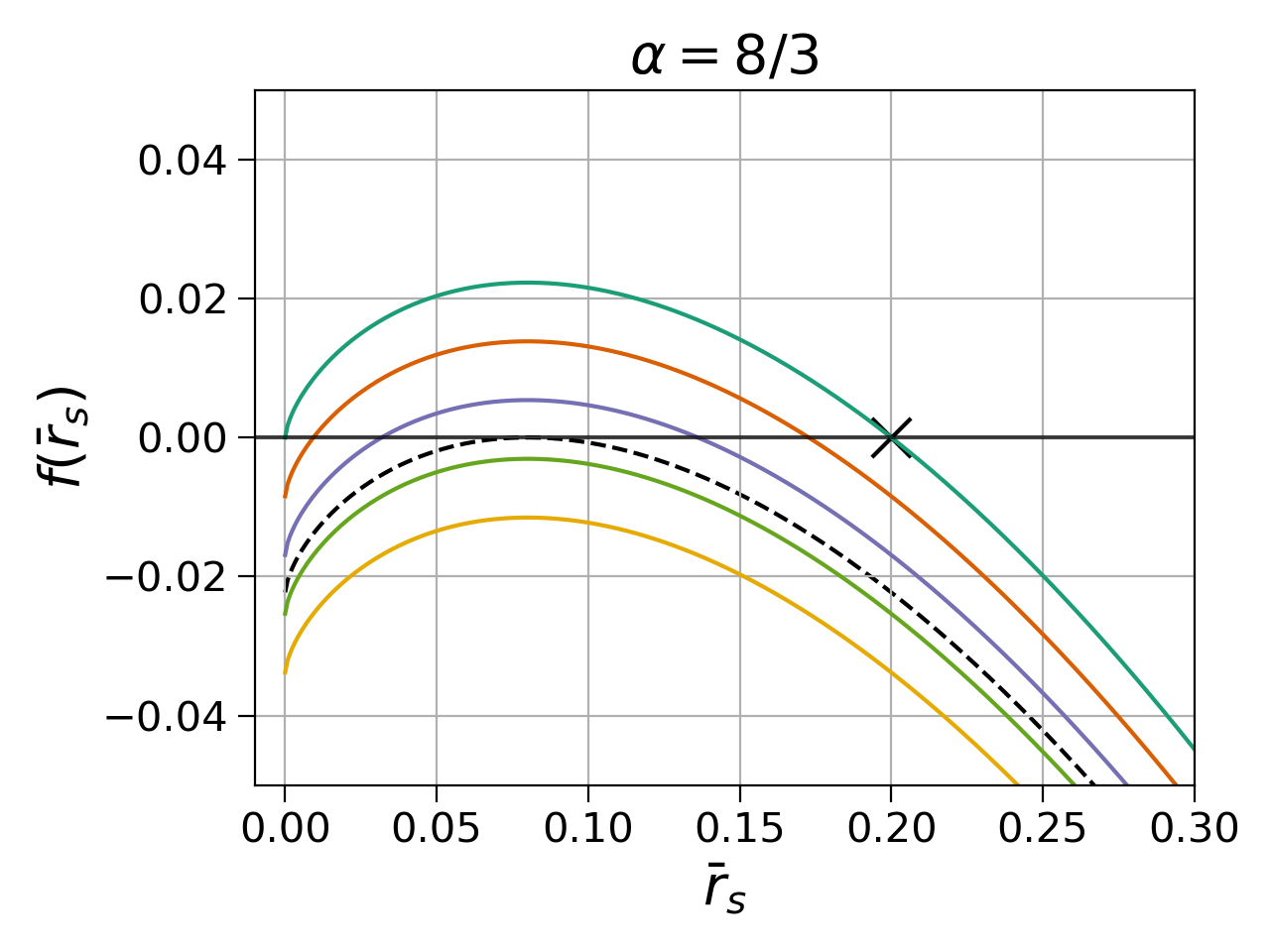}
\end{center}
\caption{Solutions for $\bar r_s$ for
  three examples for cases 1, 2 and 3.  In each panel we plot
  $f(\bar r_s)$, as defined in Eq.~(\ref{r_s}), for $\gamma = 1.4$,
  and for different values of the nondimensional heating rate
  parameter $\bar A^*$ (the labeling is the same in all three panels).  
  Solutions for $\bar r_s$ correspond to
  zero-crossings of the functions $f(\bar r_s)$.  In case 1, with
  $1 < \alpha < 2$, we find one solution for $\bar r_s$ (see the left
  panel for $\alpha = 1.5$).  In case 2, with $\alpha = 2$ (middle
  panel), we find one solution but only up to a maximum heating rate
  $\Amax = 121.6$ (marked by the dashed black line), beyond which 
  the solution $\bar r_s$ becomes negative.
  Finally, in case 3, for $\alpha > 2$, the number of solutions again
  depends on $\bar A^*$.  For the example of $\alpha = 8/3$ (shown in
  the right panel) we find two solutions for $\bar r_s$ up to a
  maximum value $\Amax = 131.8$ (marked by the dashed 
  black line), and none beyond that value.  In all three
  panels the cross marks the sonic radius $\bar r_s = 0.2$ in the
  adiabatic limit (see Eq.~\ref{r_s_adiabatic}).  In all three cases
  the (outer) sonic radius decreases as the heating rate increases.
  In case 3 an additional inner sonic radius appears for $\bar A^* >0$
  and increases for increasing heating rate, until both sonic points
  merge and disappear for the critical value $\bar A^* = \Amax$. }
\label{Fig:r_s}
\end{figure*}

As a first step we will discuss solutions for the sonic radius for
different parameter choices.  We note that
smooth and steady-state heated transonic Bondi solutions do not exist
for $\gamma = 5/3$, at least in our Newtonian treatment of the
problem.  This can be seen from Eq.~(\ref{Bernoulli_nd}), where the
first term vanishes for $\gamma = 5/3$, leaving us with
\begin{equation}
\bar r_s^{\alpha - 1} = - \beta \bar A^* \bar r_{\rm ann}^\alpha.
\end{equation}
As we discussed above, $\beta > 0$ for $\alpha > 1$, so that this
equation will not allow real and positive solutions.  We therefore
conclude that heated transonic solutions are possible only for
$\gamma < 5/3$, which we will consider in the following.  We also find
that the behavior depends on the values of $\alpha$, and we therefore
distinguish between three different cases, which are illustrated
in Fig.~\ref{Fig:r_s}.

\begin{figure*}
\begin{center}
\includegraphics[width=0.3\textwidth]{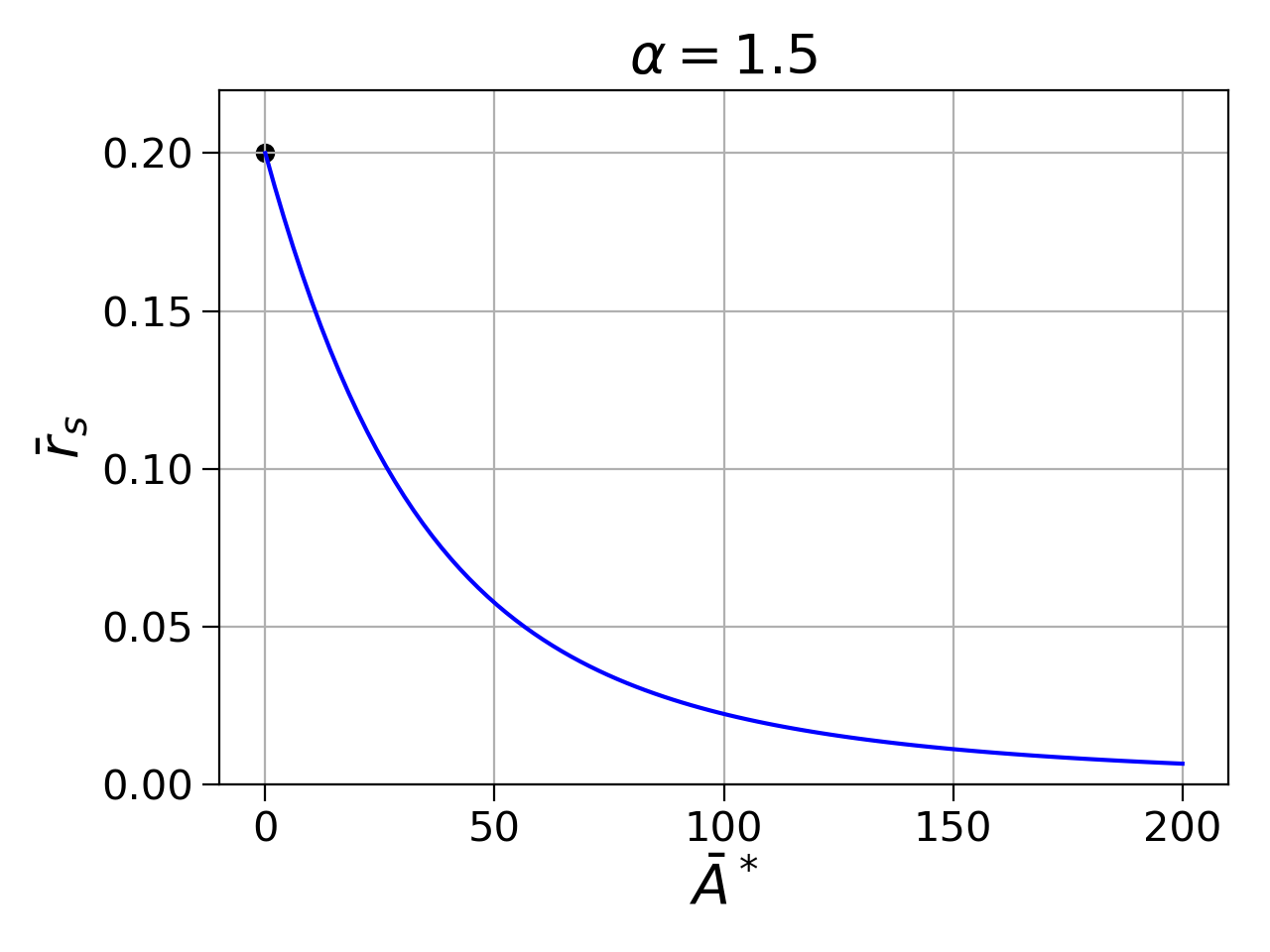}
\includegraphics[width=0.3\textwidth]{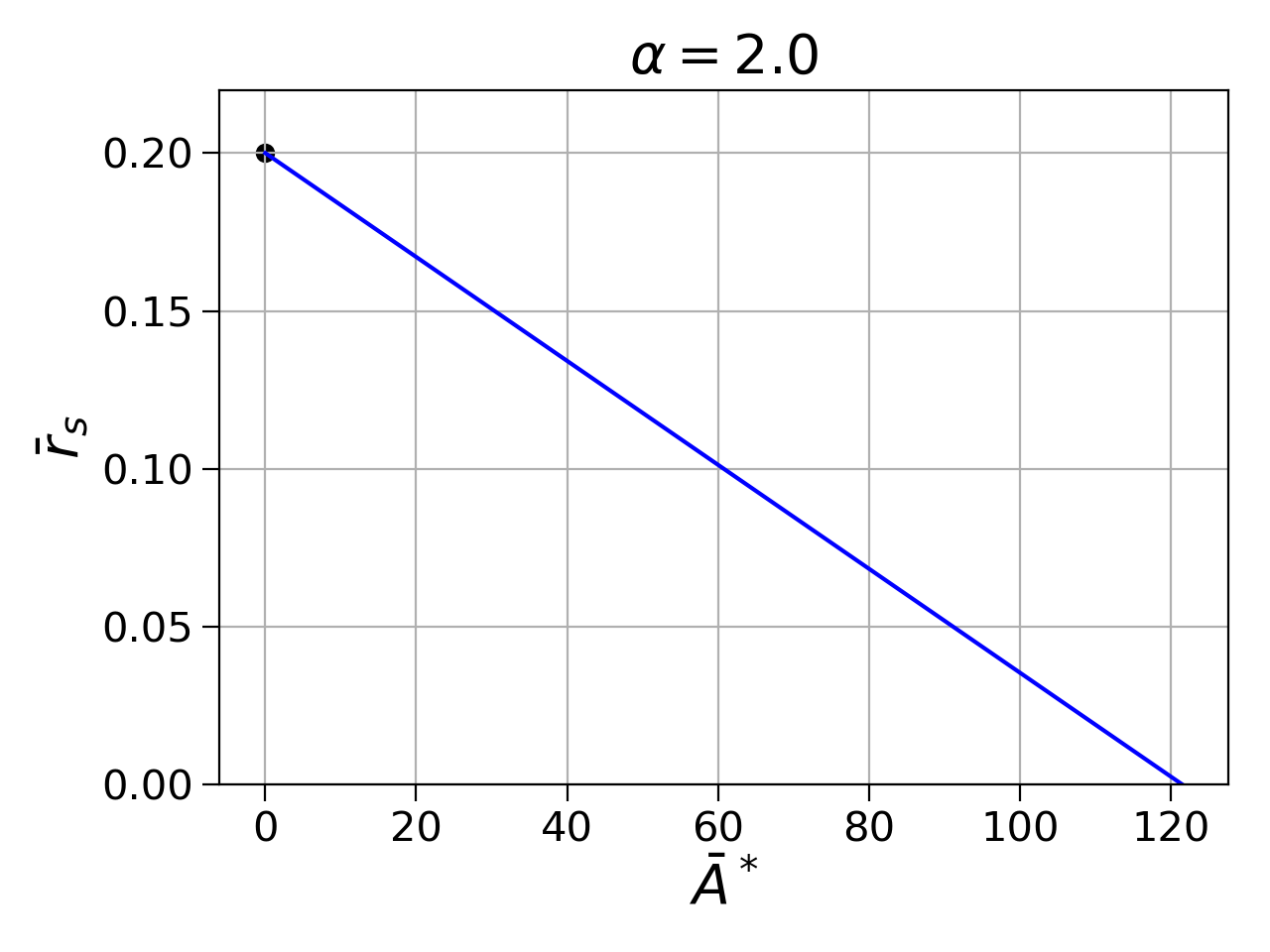}
\includegraphics[width=0.3\textwidth]{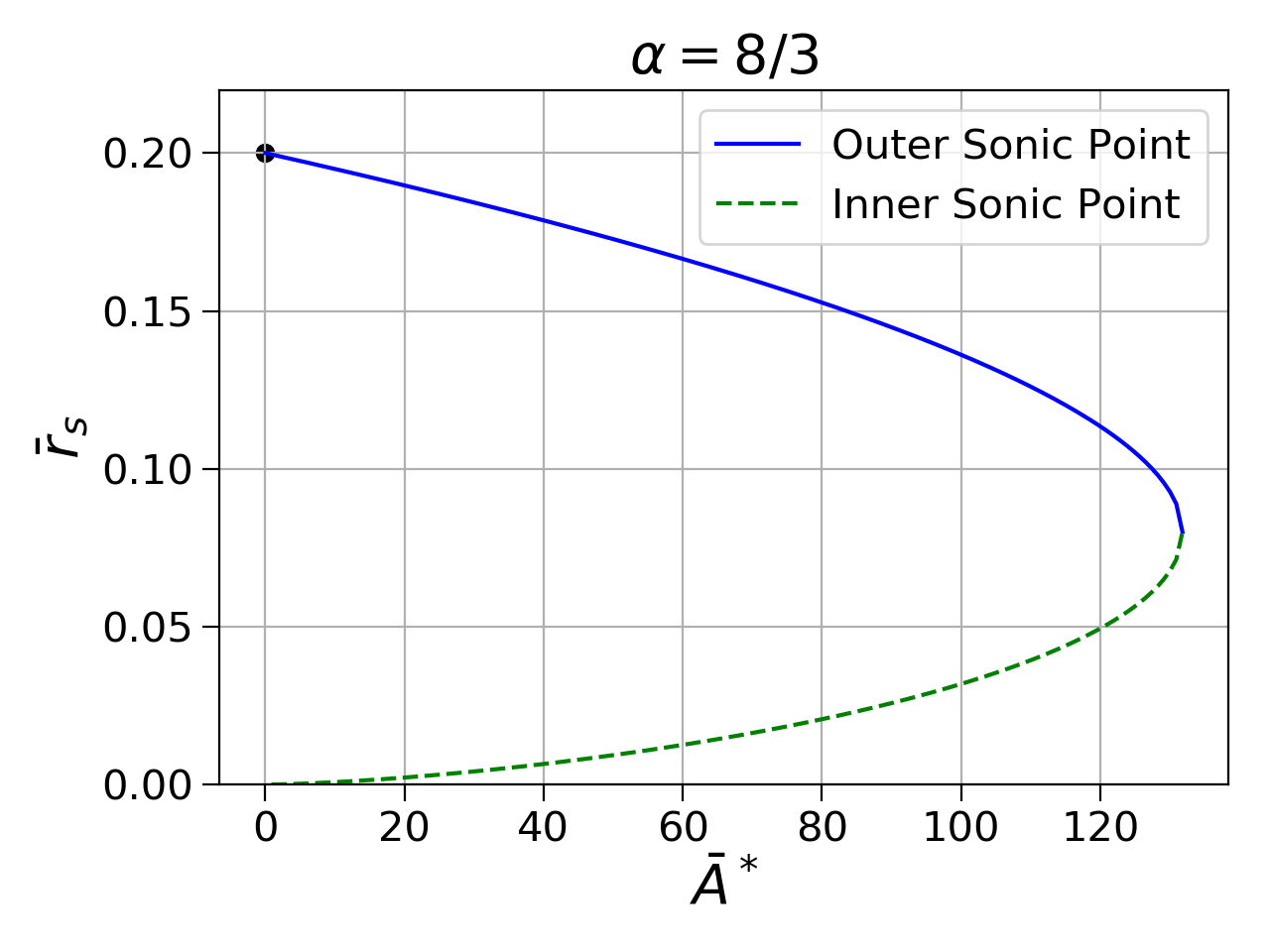}
\end{center}
\caption{Values of $\bar r_s$ as a function of $\bar A^*$ for the same
  three examples as shown in Fig.~\ref{Fig:r_s}, namely for $\alpha =
  1.5$ (left panel), $\alpha = 2$ (middle panel) and $\alpha = 8/3$
  (right panel), and all for $\gamma = 1.4$.  Note that (i) a solution for
  $\bar r_s$ exists for all values of $\bar A^*$ for $\alpha < 2$
  (case 1; left panel), (ii) the linear behavior for $\alpha = 2$ (case 2;
  middle panel), and (iii) the existence of two solutions for $\bar A^* <
  \Amax$ for $\alpha = 8/3$ (case 3; right panel).}
\label{Fig:r_s_2}
\end{figure*}

%===============================================================================
\subsubsection{Case 1: $1 < \alpha < 2$}
%===============================================================================

In the regime $1 < \alpha < 2$ we find one single real value for $\bar
r_s$ for suitable combinations of $\bar A^*$ and $\gamma < 5/3$.  An
example, for $\gamma = 1.4$, is shown in the left panel of
Fig.~\ref{Fig:r_s}, where the cross denotes the sonic radius $\bar r_s
= 0.2$ in the adiabatic limit (see Eq.~\ref{r_s_adiabatic}).  Note
that the sonic radius decreases with increasing heating parameter $\bar
A^*$, indicating that heating prevents the flow from becoming transonic
until it gets closer to the black
hole.  We also show $\bar r_s$ as a function of $\bar A^*$ in the left
panel of Fig.~\ref{Fig:r_s_2}.

%===============================================================================
\subsubsection{Case 2: $\alpha = 2$}
\label{sec:find_r_s_equal_2}
%===============================================================================

In the special case of $\alpha = 2$, Eq.~(\ref{r_s}) reduces to the
linear equation
\begin{equation} \label{r_s_2}
\bar r_s = \frac{5 - 3 \gamma}{4} - \beta \bar A^* \bar r_{\rm ann}^2 \hfill (\alpha = 2)
\end{equation}
providing us with a unique value of $\bar r_s$ (see also the middle
panels in Figs.~\ref{Fig:r_s} and \ref{Fig:r_s_2}).  Evidently, we can
find positive solutions for $\bar r_s$ only for
\begin{equation}
\bar A^* < \Amax = \frac{5 - 3 \gamma}{4} \frac{1}{\beta \bar r_{\rm ann}^2}.
\hfill (\alpha = 2)
\end{equation}
In particular, we find $\Amax = 0$ for $\gamma = 5/3$,
consistent with our discussion above.  As in case 1, increasing the
heating rate will decrease the sonic radius.

%===============================================================================
\subsubsection{Case 3: $\alpha > 2$}
\label{sec:find_r_s_greater_2}
%===============================================================================

An example for the case $\alpha > 2$, for $\alpha = 8/3$, is shown in
the right panel of Fig.~\ref{Fig:r_s}.  The cross again marks the
sonic radius $\bar r_s = 0.2$ in the adiabatic limit with
$\bar A^* = 0$.  For $\bar A^* > 0$ an inner sonic point emerges,
suggesting that, as the gas accretes, it becomes supersonic at the
outer sonic point, but does not remain supersonic.

%This behavior can be anticipated from Eq.~\ref{u_prime_nd}.  For
%$\alpha < 2$, the gravitational term $1/\bar r^2$ in $\bar D_1$ will
%dominate the numerator of the right-hand side at small $\bar r$,
%ensuring that $\bar u' < 0$, so that $\bar u$ continues to increase as
%the gas accretes onto the black hole.  For $\alpha > 2$, however, the
%heating term $\bar H$ will dominate for sufficiently small $\bar r$.
%Since this term enters with the opposite sign, the gas speed will take
%a maximum, and will then decrease inside this maximum, suggesting that
%it will become subsonic again.  For sufficiently large $\bar A^*$ the
%gas speed's turning point will occur before the gas becomes
%supersonic, so that no transonic solution exists.  In the right panel
%of Fig.~\ref{Fig:r_s_2} we show the two transonic points as a function
%of $\bar A^*$; as we anticipated, solutions for $\bar r_s$ exist only
%up to a maximum heating rate $\Amax$.

To find the critical value $\Amax$ above which no transonic radius
exists we consider Eq.~(\ref{r_s}) an equation for $\bar A^*$ as a
function of $\bar r_s$ rather than the other way around (effectively
flipping the axes in the right panel of Fig.~\ref{Fig:r_s_2}).  The
critical value $\Amax$ is then given by the point at which the
derivative $d \bar A^* / d \bar r_s$ vanishes, which yields
\begin{equation}
\rsmax = \frac{\alpha -2}{\alpha -1} \frac{5 - 3 \gamma}{4}.
\hfill (\alpha > 2)
\end{equation}
Inserting this into (\ref{r_s}) and solving for $\Amax$ yields
\begin{equation}
\Amax = \frac{1}{\alpha - 1} \frac{(\rsmax)^{\alpha - 1}}{\beta} 
\frac{1}{\bar r_{\rm ann}^\alpha}.
\hfill (\alpha > 2)
\end{equation}
We again find that $\Amax = 0$ for $\gamma = 5/3$, consistent with our
discussion above.  For other suitable values of $\gamma < 5/3$ and
$\bar A^* < \Amax$, however, we find two solutions for the sonic
radius $\bar r_s$.

\begin{figure*}
\begin{center}
\includegraphics[width=0.3\textwidth]{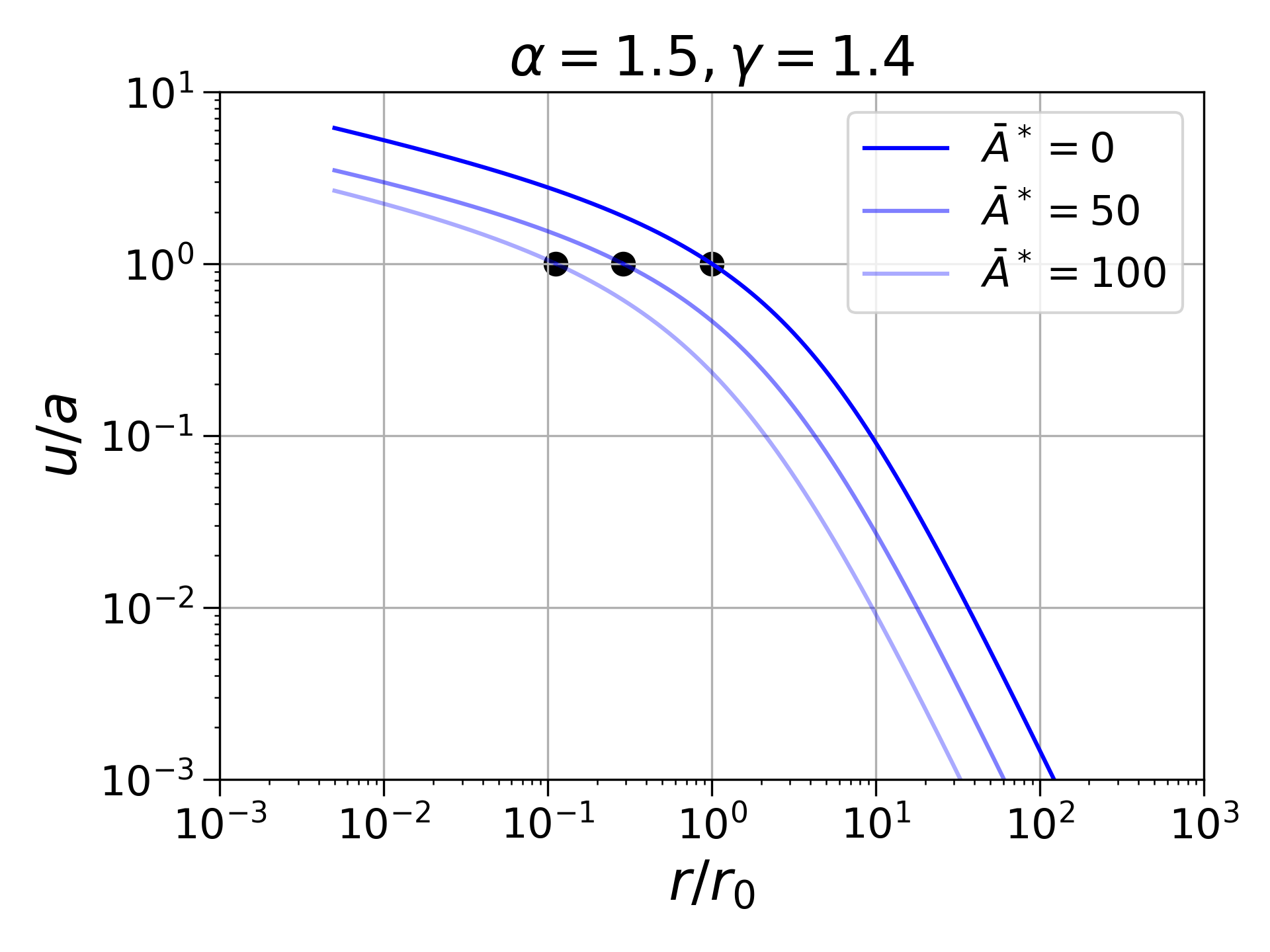}
\includegraphics[width=0.3\textwidth]{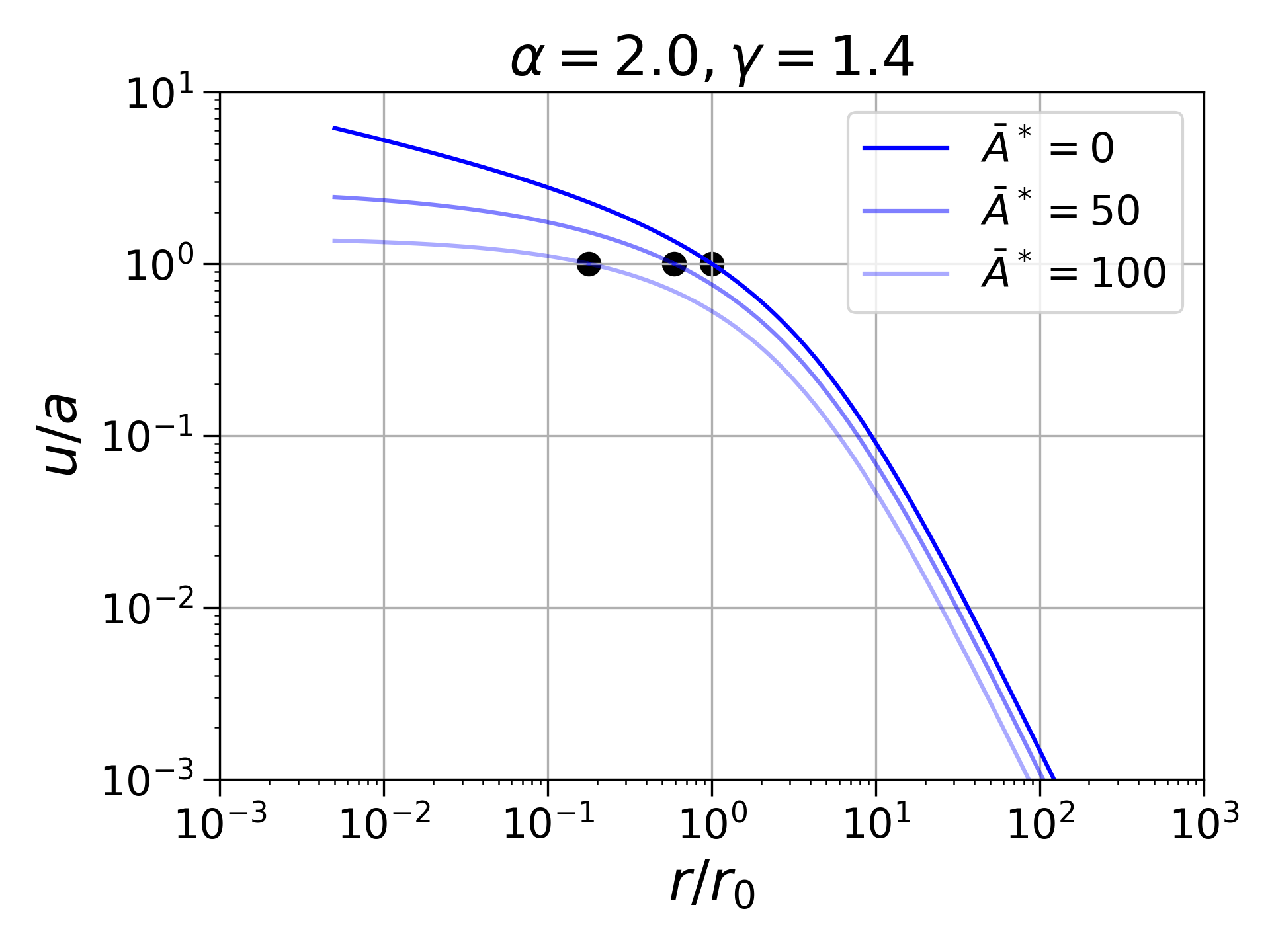}
\includegraphics[width=0.3\textwidth]{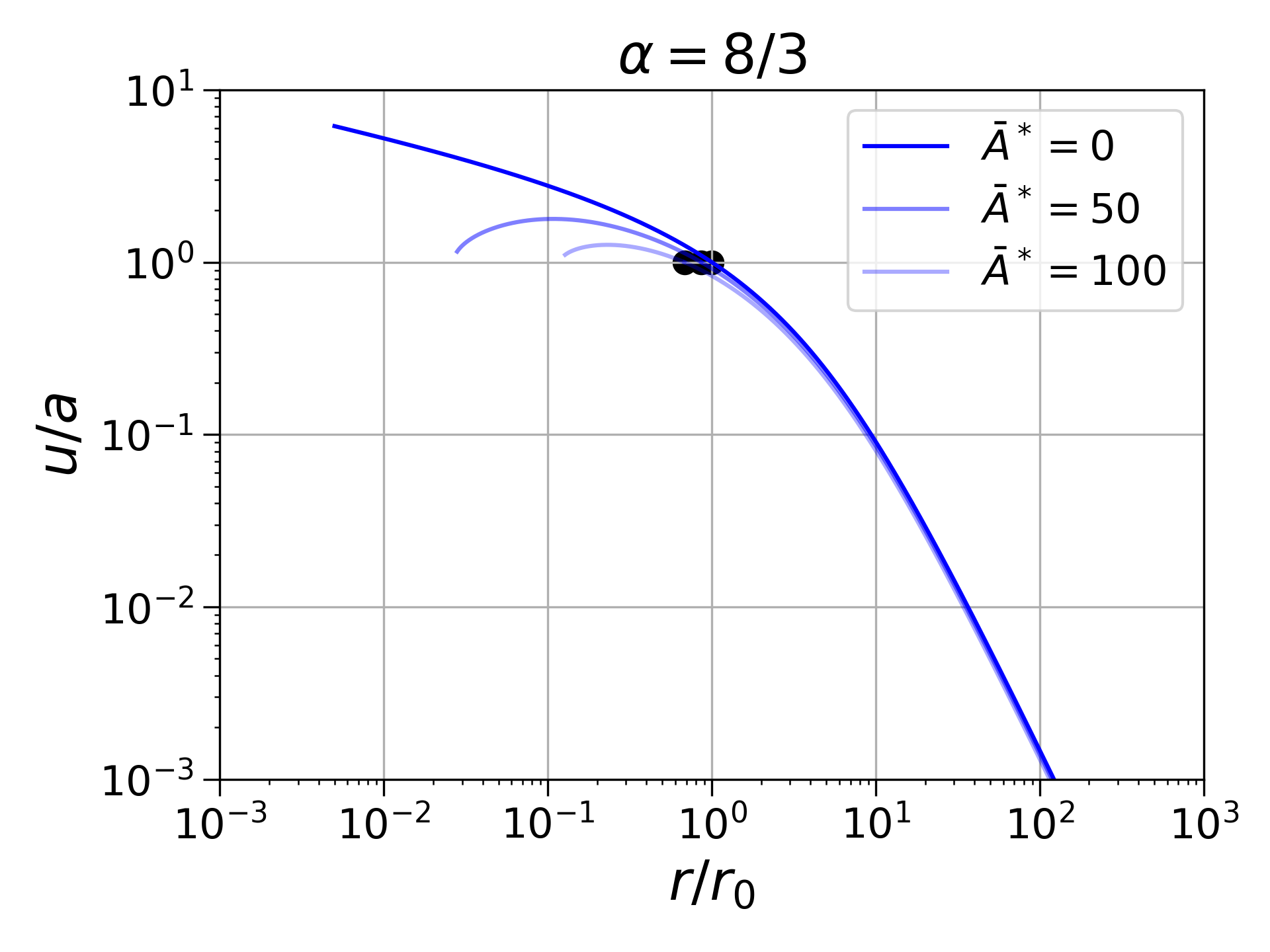}
\end{center}
\caption{Examples of fluid profiles with $\gamma = 1.4$ for the same 
three examples as shown in Figs.~\ref{Fig:r_s} and \ref{Fig:r_s_2}.  
The left panel shows the fluid velocity $u$ for $\alpha = 1.5$, the middle 
panel for $\alpha = 2.0$, and the right panel for $\alpha = 8/3$.  The (outer) 
sonic points, at which $u = a$, are marked by dots.  For $\alpha = 2.0$, the 
sonic radius $\bar r_s$ goes to zero as $\bar A^*$ approaches $\Amax$.  
For $\alpha = 8/3$, the fluid approaches a singular inner sonic point.}
\label{Fig:u}
\end{figure*}

\begin{figure*}
\begin{center}
\includegraphics[width=0.3\textwidth]{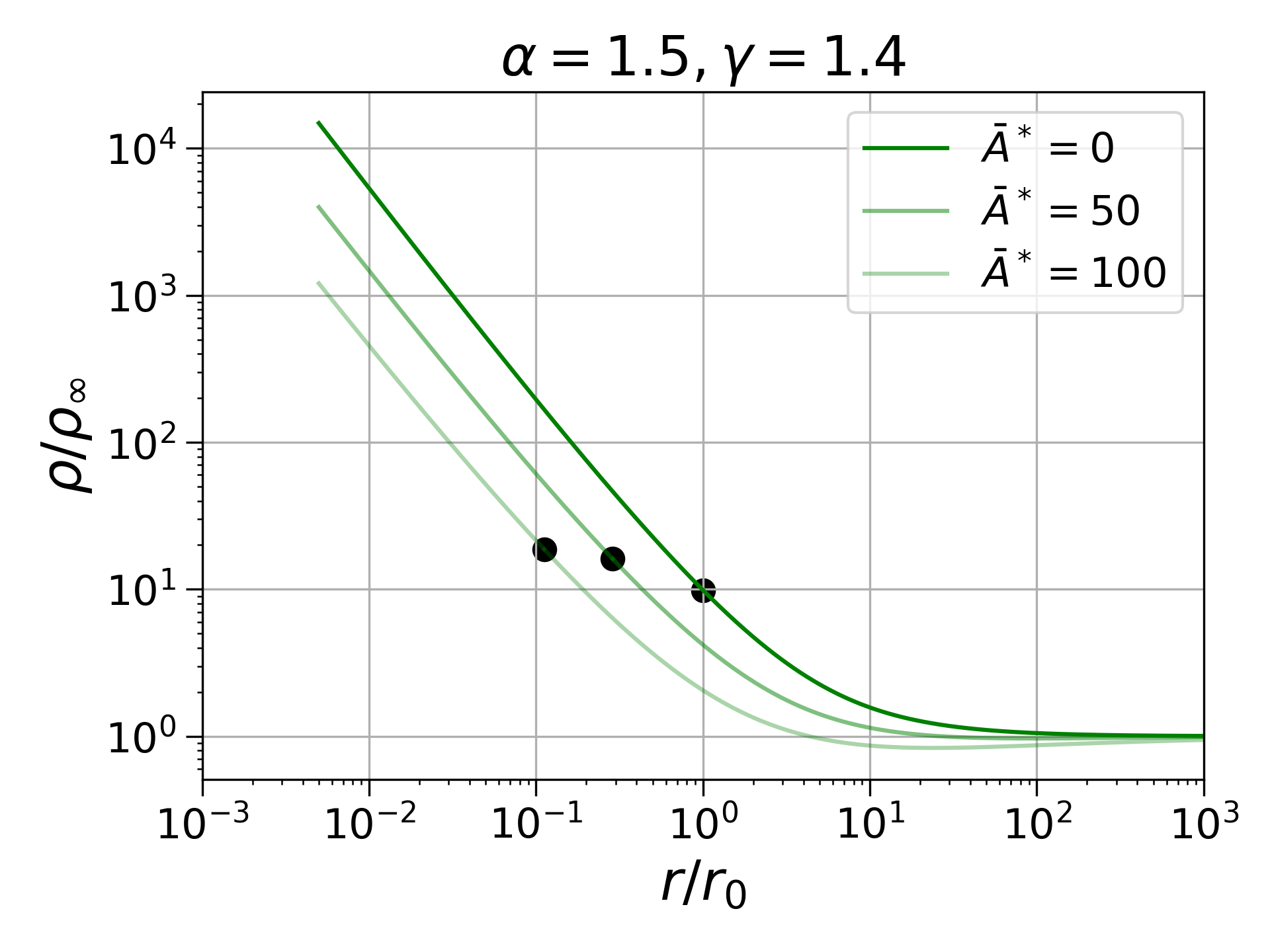}
\includegraphics[width=0.3\textwidth]{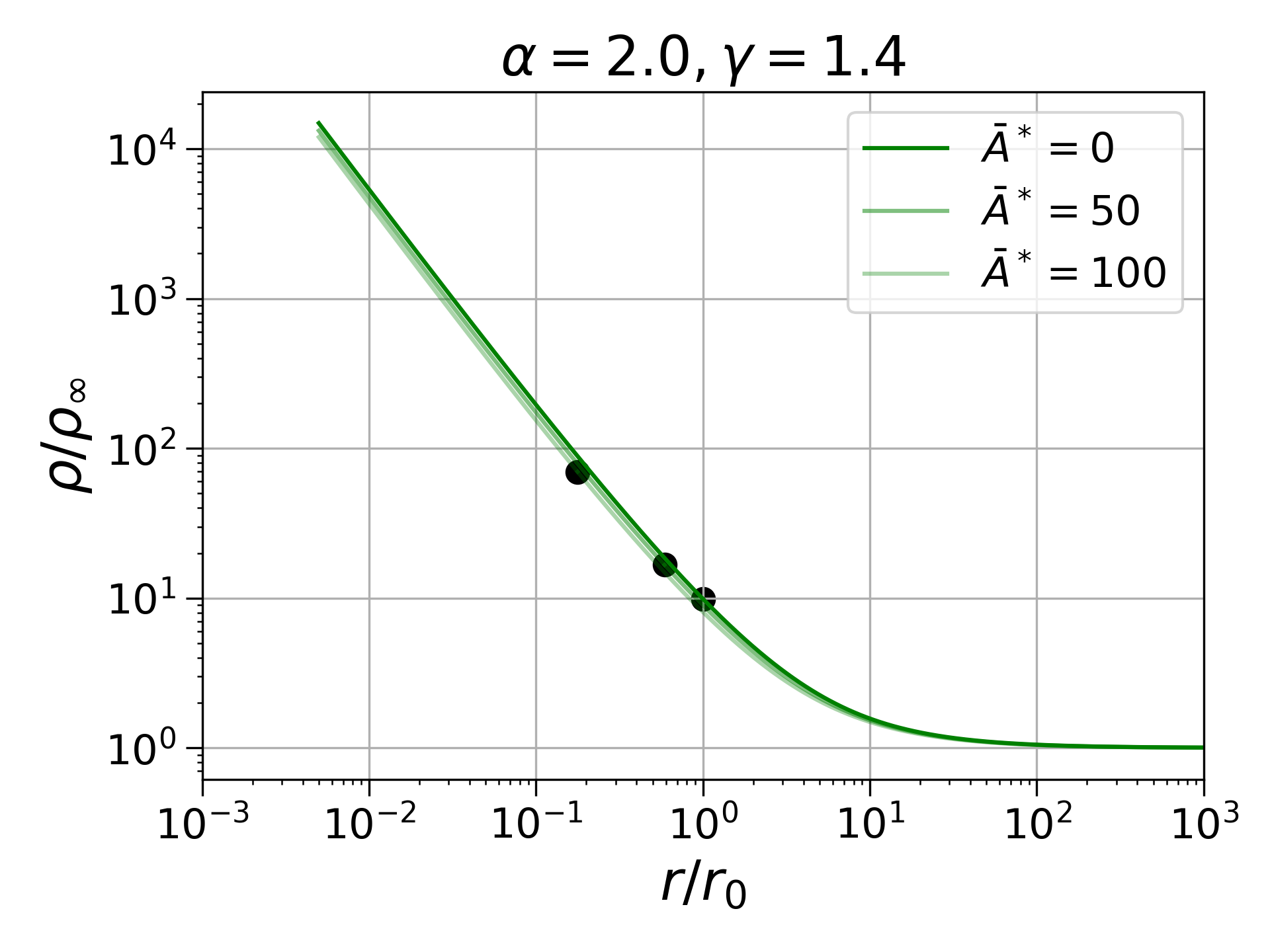}
\includegraphics[width=0.3\textwidth]{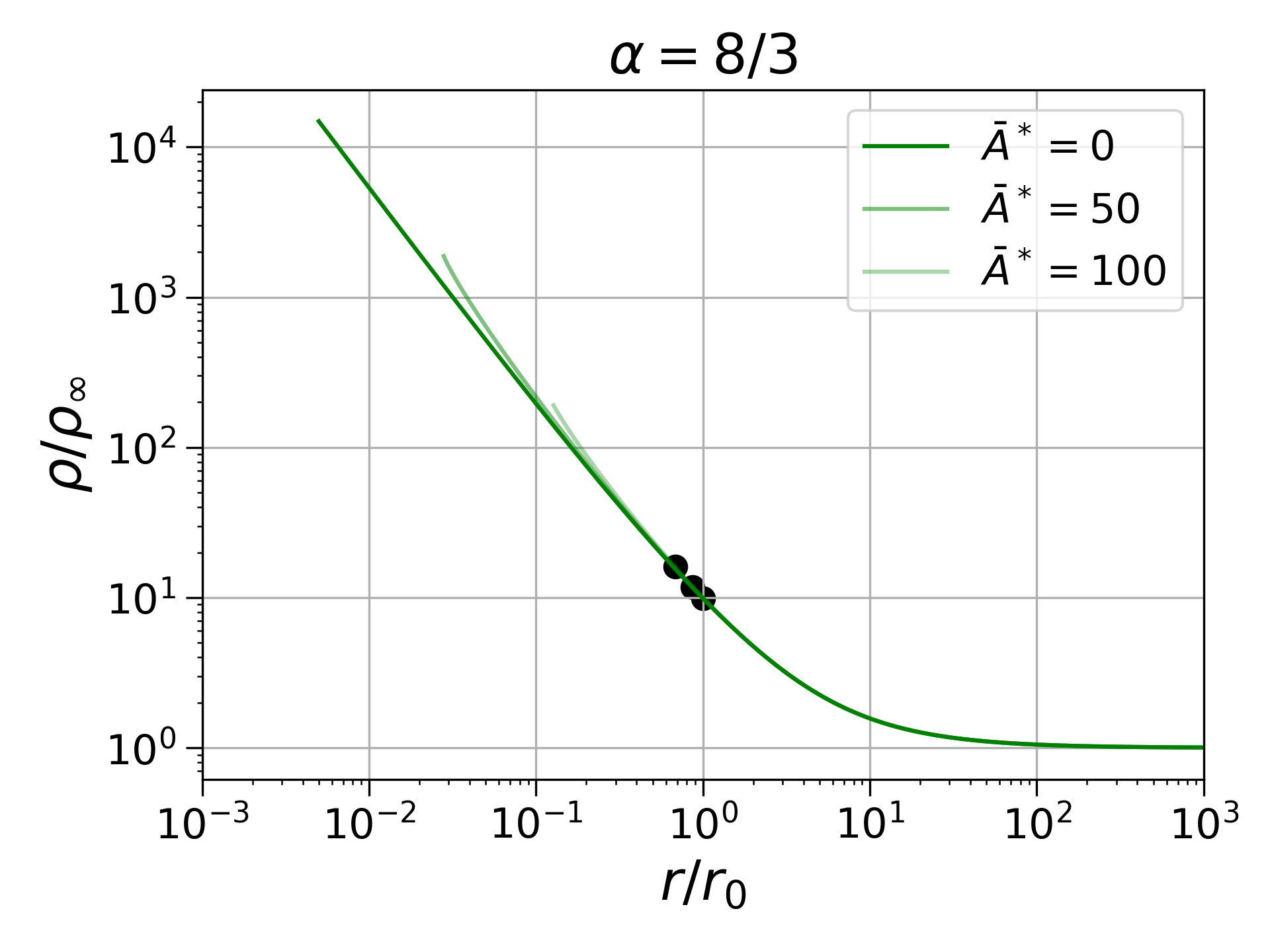}
\end{center}
\caption{Same as Fig.~\ref{Fig:u}, but for density profiles.}
\label{Fig:rho}
\end{figure*}

As described in Section \ref{sec:transonic_strategy},
constructing fluid flow profiles requires an expansion about the sonic
radii $\bar r_s$, since the differential equations (\ref{u_prime_nd})
and (\ref{rho_prime_nd}) cannot be evaluated directly at those points.
Applying l'H\^opital's rule results in a quadratic
equation for $\bar u'$.  In all cases that we have considered, this
equation had real solutions at the outer sonic point, allowing for
smooth flow there, but only imaginary solutions at the inner sonic
point.  This is an indication that it is impossible to construct
smooth solutions across the inner sonic point, where the fluid's speed
drops from being supersonic to subsonic.  Instead, we might expect
that shocks, and hence discontinuities in the fluid's flow,
develop at this point \citep[see also][]{ChaO85,ParO98}.  As a
result, we conclude that in the regime considered, for $\alpha > 2$,
no smooth, steady-state transonic solutions describing spherical
accretion exist.

For $\alpha \geq 3$ we might find even more
solutions for $\bar r_s$, but we do not pursue this possibility in
greater detail, since this range of parameters appears
less relevant astrophysically.

%===============================================================================
\subsection{Finding fluid profiles and the accretion rate}
\label{sec:find_M_dot}
%===============================================================================

As outlined in Section \ref{sec:transonic_strategy}, finding the
accretion rate involves an iterative ``shooting method" to match to
the boundary conditions at $\bar r \rightarrow \infty$.  This involves
integrating the differential equations (\ref{u_prime_nd}),
(\ref{rho_prime_nd}) and (\ref{K_prime_nd}), which, in turn, involves
applying l'H\^opital's rule at the (outer) sonic radius $\bar r_s$.   Once 
$\bar K_s$ has been found, the equations can be integrated both 
outwards and inwards in order to find the profiles of the fluid flow.  We show 
examples for the three different cases with $\alpha < 2$, $\alpha = 2$ and 
$\alpha > 2$ in Figs.~\ref{Fig:u} and \ref{Fig:rho}.

\begin{figure}
\begin{center}
\includegraphics[width=0.45\textwidth]{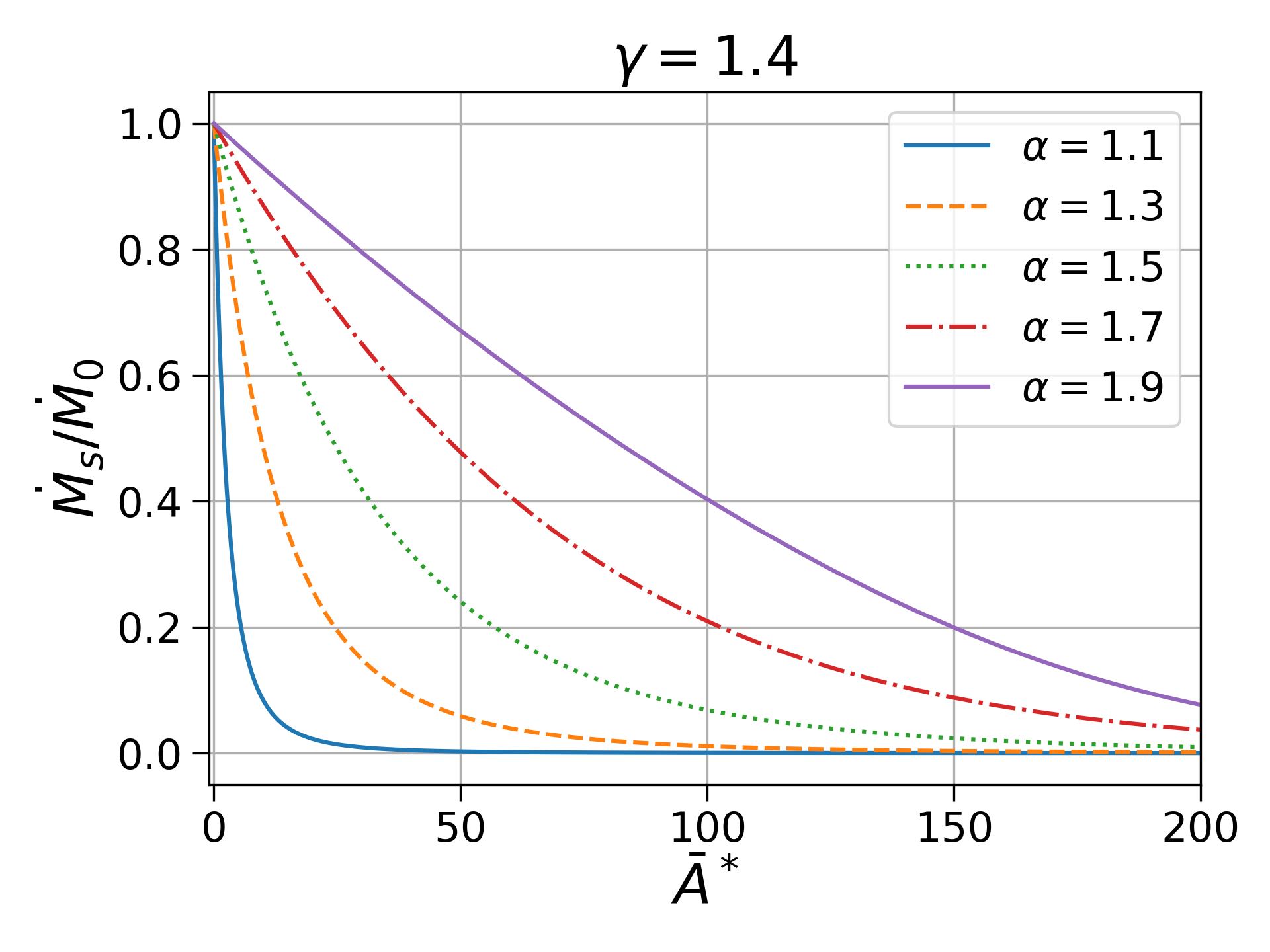}
\includegraphics[width=0.45\textwidth]{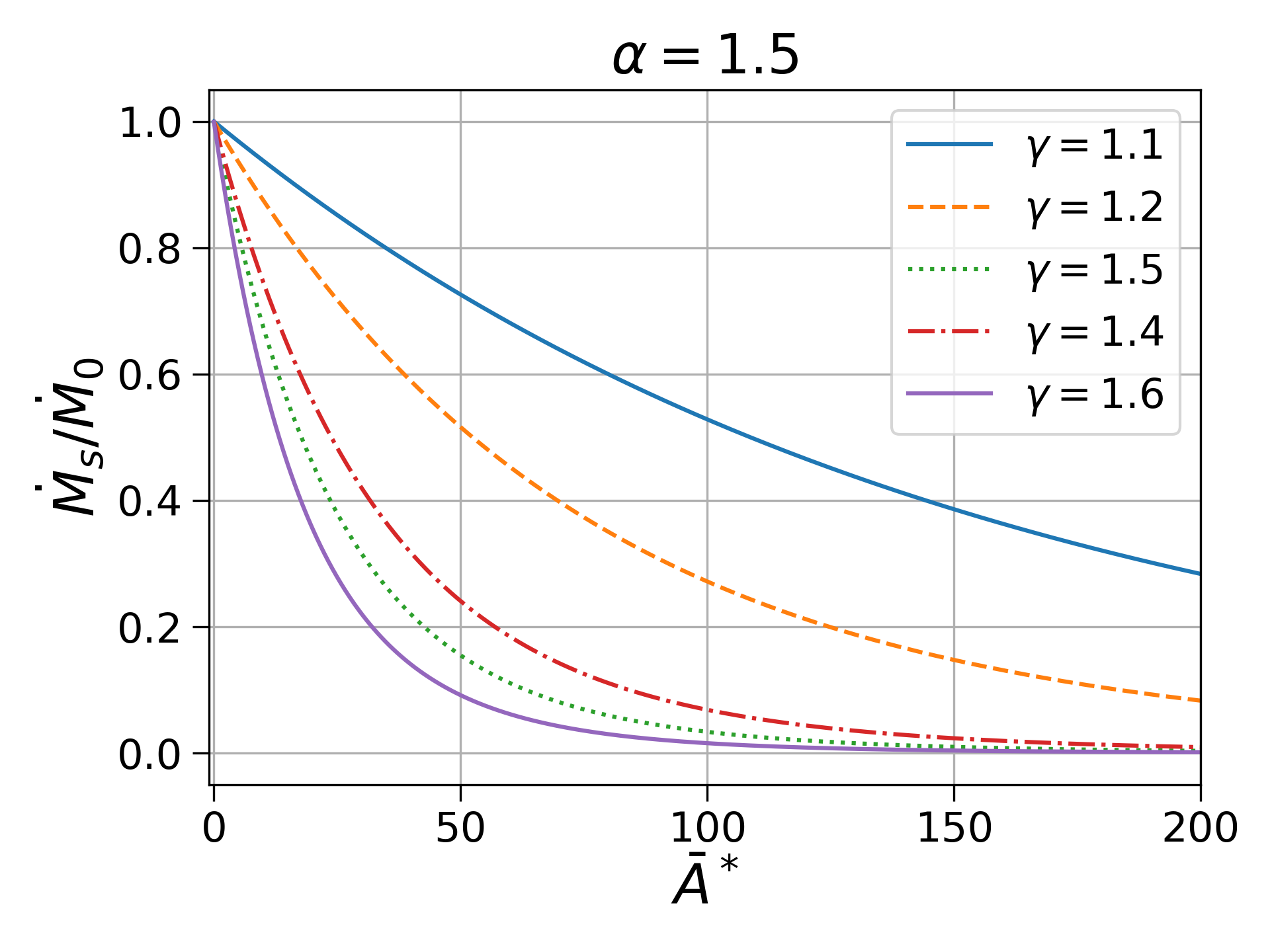}
\end{center}
\caption{The accretion rate $\dot M_s / \dot M_0$ for transonic flow as a function of $\bar A^*$ for different sets of parameters. The upper panel shows results for fixed $\gamma = 1.4$ and varying values of $1 < \alpha < 2$ while the lower panel shows results for a fixed $\alpha = 1.5$ and varying values of $1 < \gamma < 5/3$. As expected, the accretion rate decreases with increasing heating parameter $\bar A^*$. We find that the accretion rate decreases more rapidly as the DM power law heating $\alpha$ approaches $1$ and the adiabatic index $\gamma$ approaches $5/3$. }
\label{Fig:M_dot}
\end{figure}

As we discussed above, $\alpha > 2$ leads to the existence of a second inner sonic point, across which we cannot find smooth solutions (see also the right panel in Fig.~\ref{Fig:u}).  We therefore focus on $\alpha < 2$ here.  We show examples for different values of $1 < \alpha < 2$ and $\gamma = 1.4$, which we have previously considered in the left panels of Figs.~\ref{Fig:r_s} through \ref{Fig:rho}, and show a graph of $\dot M / \dot M_0$ as a function of $\bar A^*$ in the upper panel of Fig.~\ref{Fig:M_dot}. We also show a graph for $\dot M / \dot M_0$ as a function of $\bar A^*$ for different values of $1 < \gamma < 5/3$ and $\alpha = 1.5$ in the lower panel of Fig.~\ref{Fig:M_dot}. As anticipated, the heating due to DM annihilation reduces the accretion rate.

%===============================================================================
\section{Heated subsonic flow}
\label{sec:subsonic_results}
%===============================================================================

\begin{figure*}
\begin{center}
\includegraphics[width=0.3\textwidth]{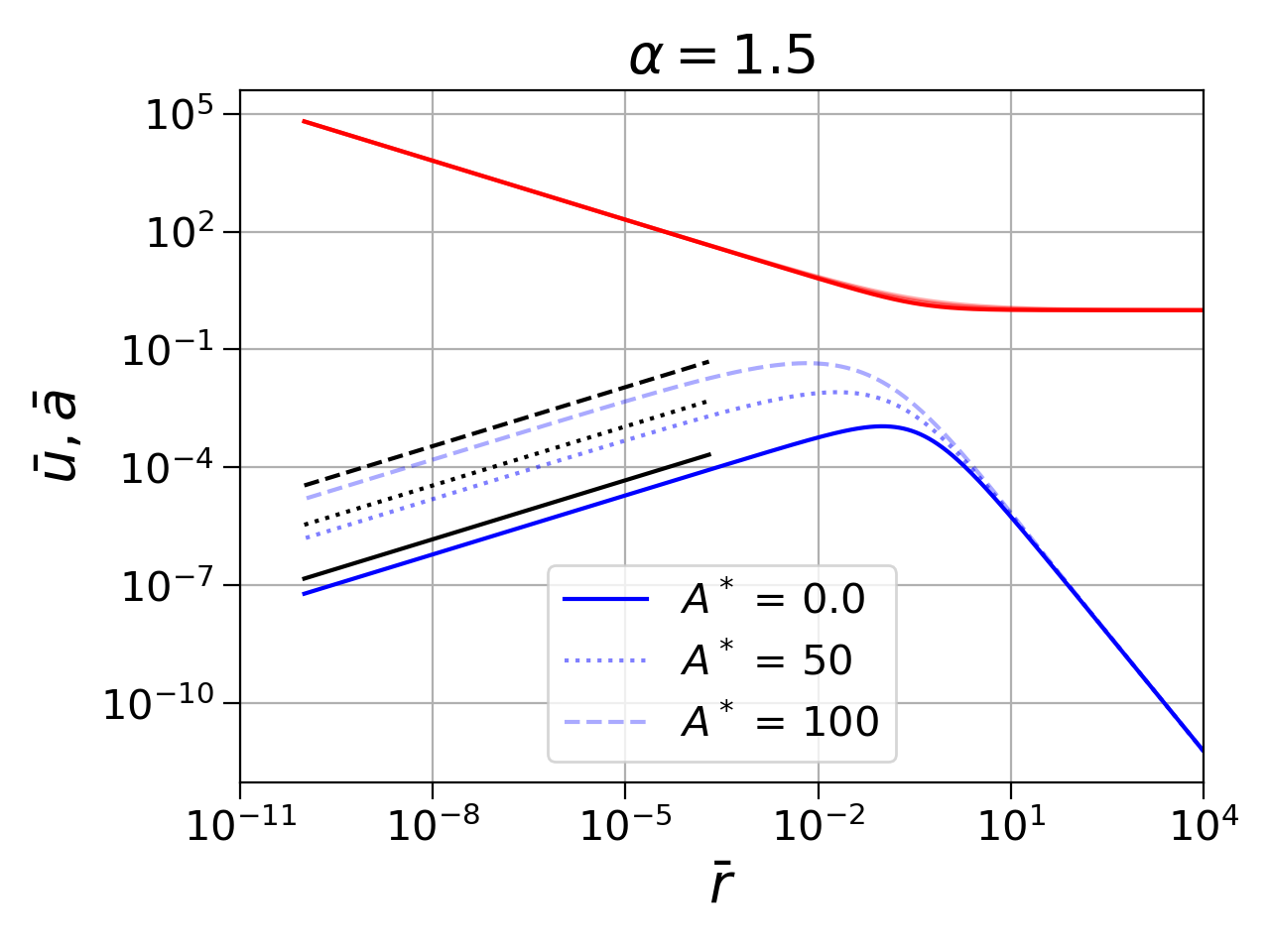}
\includegraphics[width=0.3\textwidth]{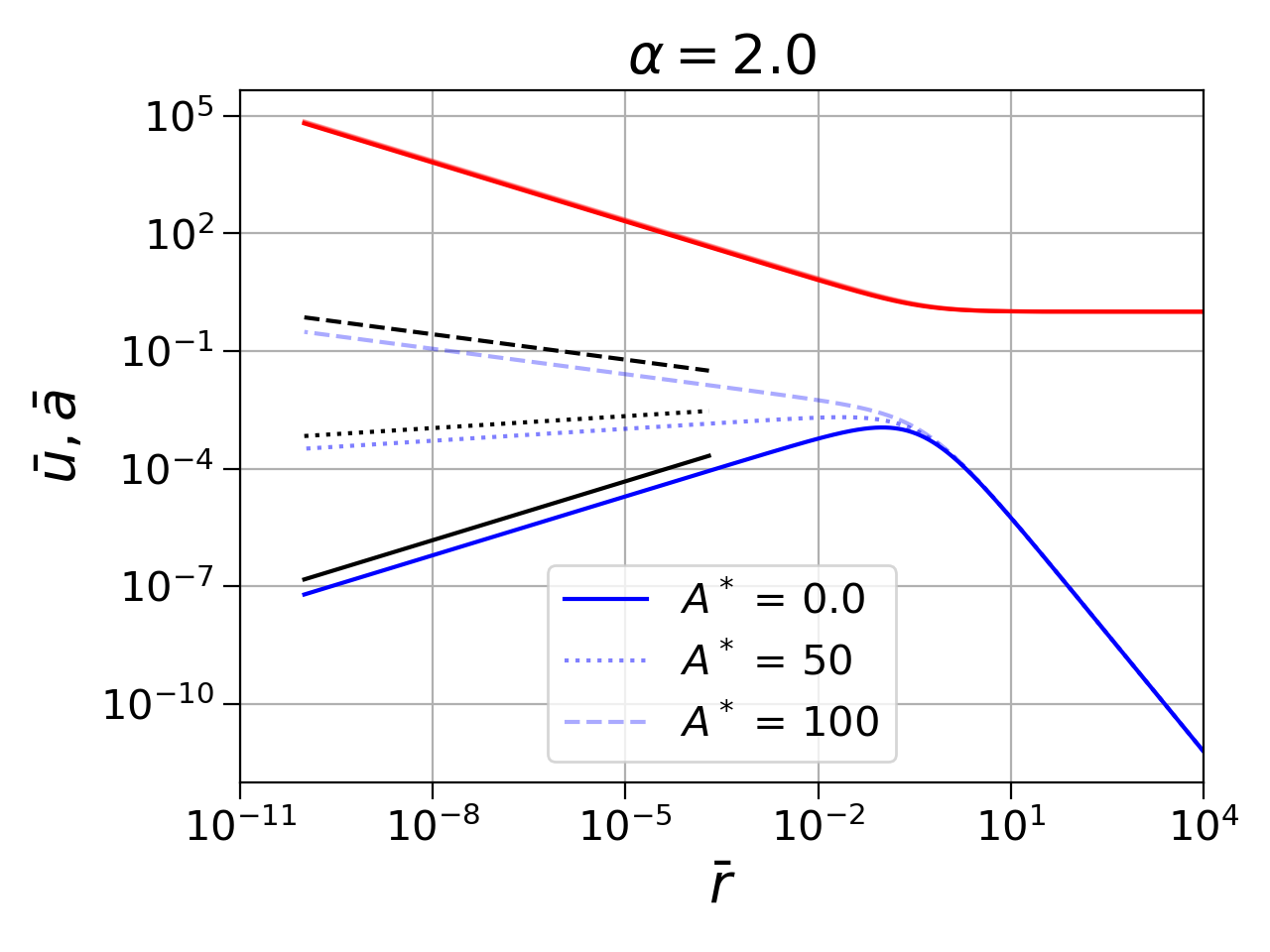}
\includegraphics[width=0.3\textwidth]{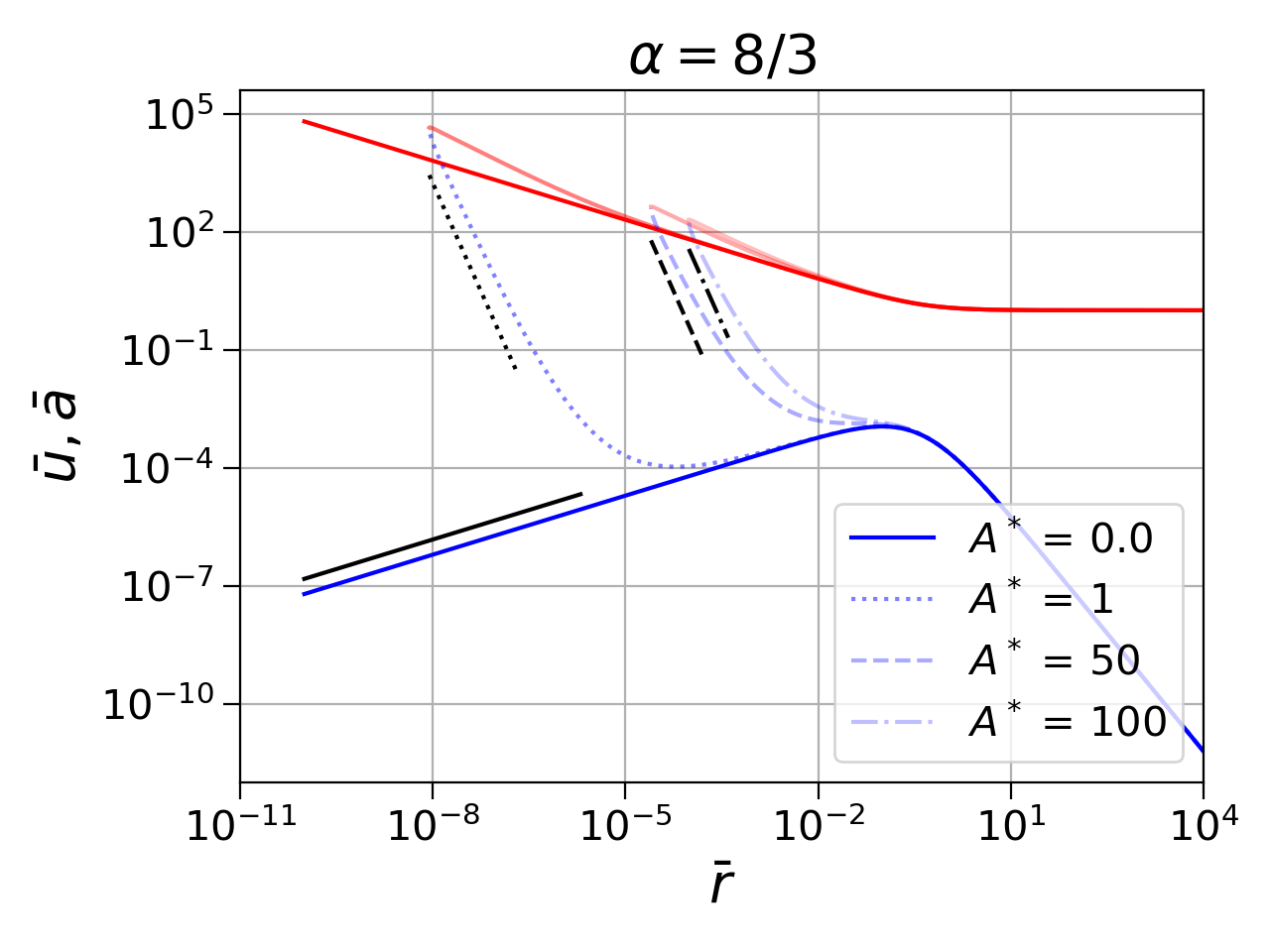}
\end{center}
\caption{Subsonic flow profiles for $\gamma = 1.4$, with $\alpha =
  1.5$ in the left panel, $\alpha = 2$ in the middle panel, and
  $\alpha = 8/3$ in the right panel.  We show results for selected
  values of $\bar A^*$ in all three cases, with the lower (blue) lines
  showing the fluid velocity $u$, and the upper (red) lines showing
  the sound speed $a$.  Also included are the expected power laws for
  the fluid velocity $u$, as given by Eqs.~(\ref{u_sub_1}),
  (\ref{u_sub_2}) and (\ref{u_sub_3}).  In case 1 (left panel, Section
  \ref{sec:subsonic_1}) the scaling of the fluid profiles approach the same
  power laws as their adiabatic counterparts, in case 2 (middle panel,
  Section \ref{sec:subsonic_2}) the power law exponent depends on the
  heating rate, and in case 3 (left panel, Section
  \ref{sec:subsonic_3}) heating results in the appearance of a
  singular sonic point at small radii.}
\label{Fig:subsonic}
\end{figure*}

\begin{figure*}
\begin{center}
\includegraphics[width=0.3\textwidth]{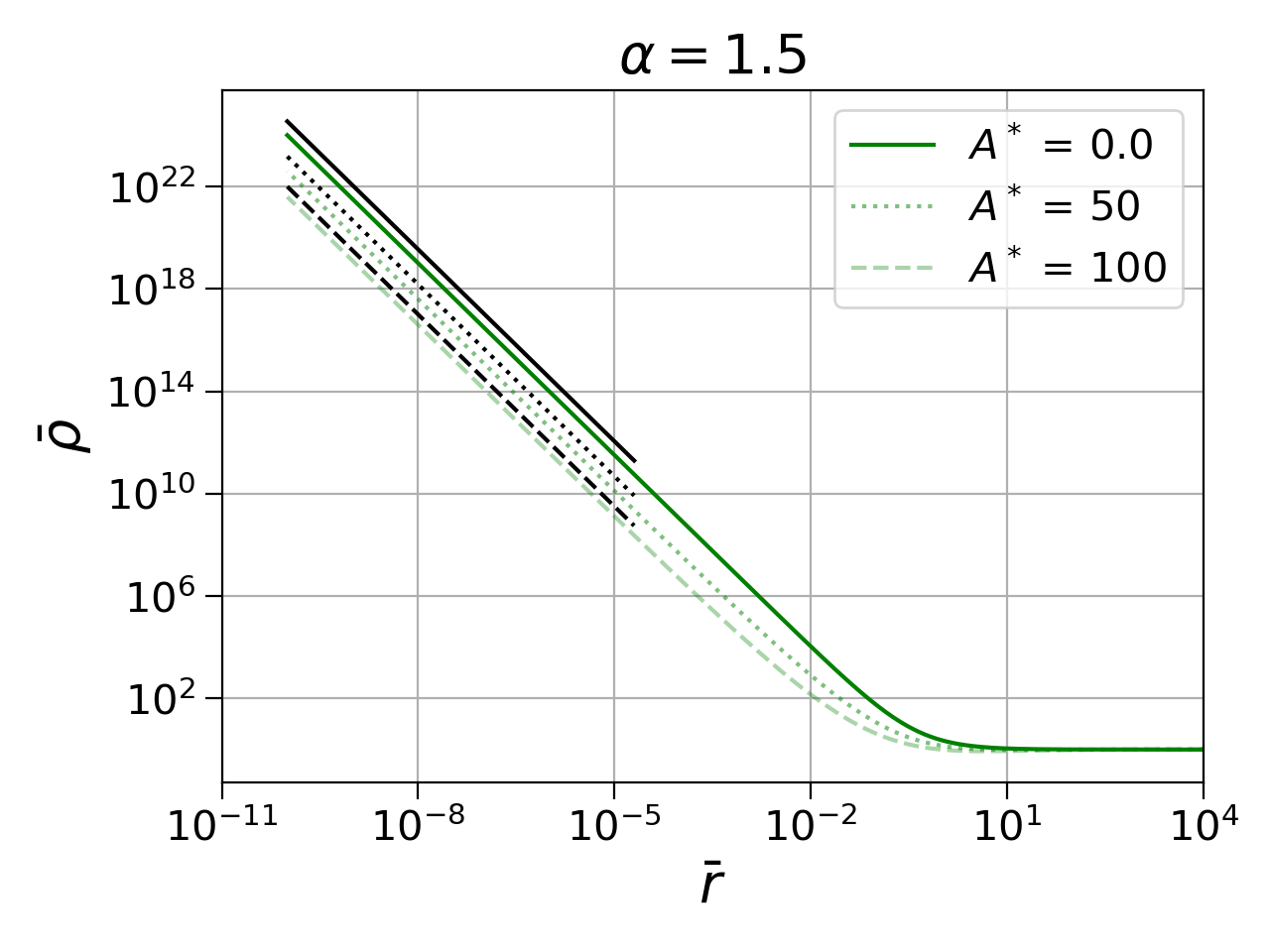}
\includegraphics[width=0.3\textwidth]{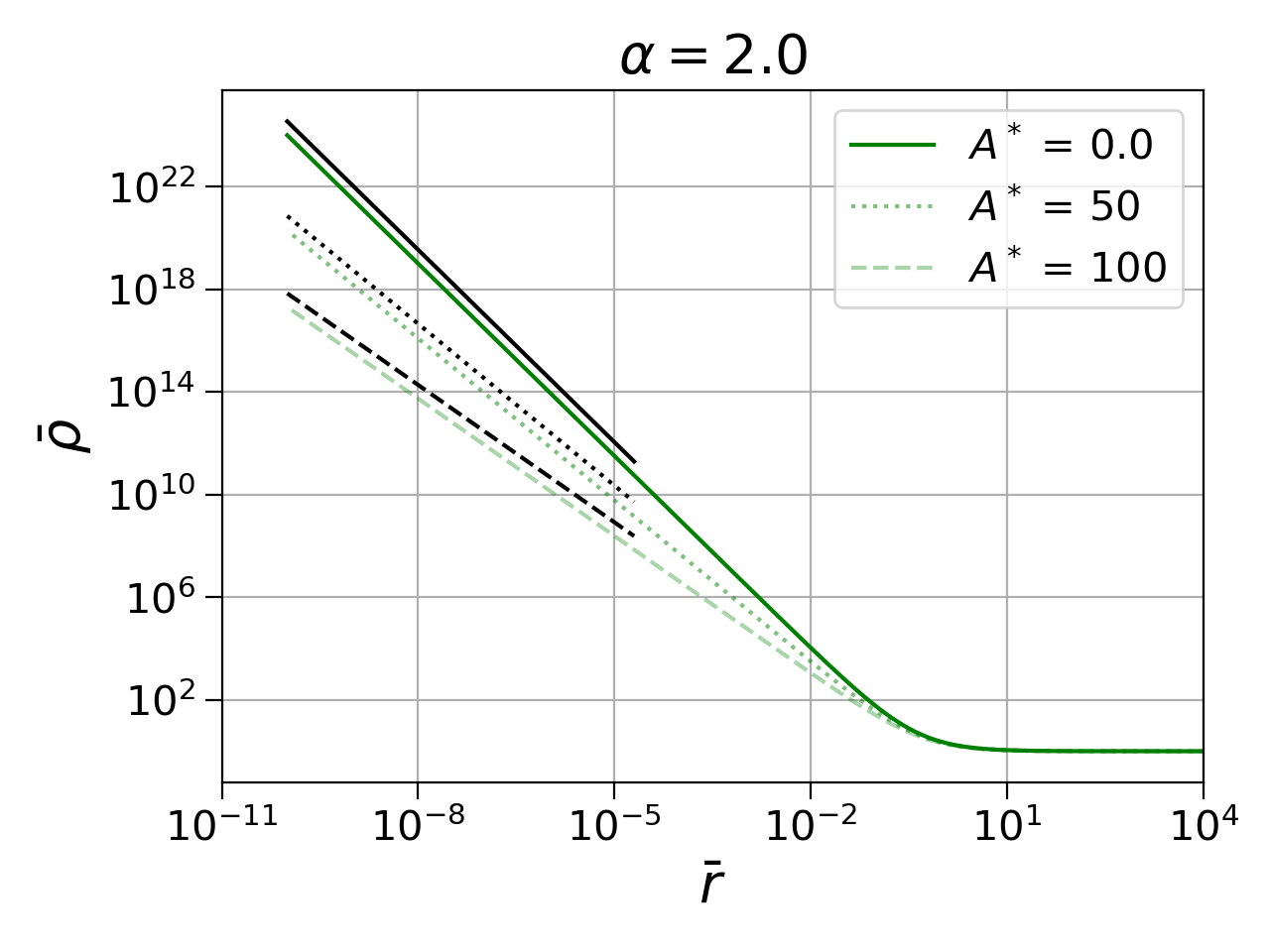}
\includegraphics[width=0.3\textwidth]{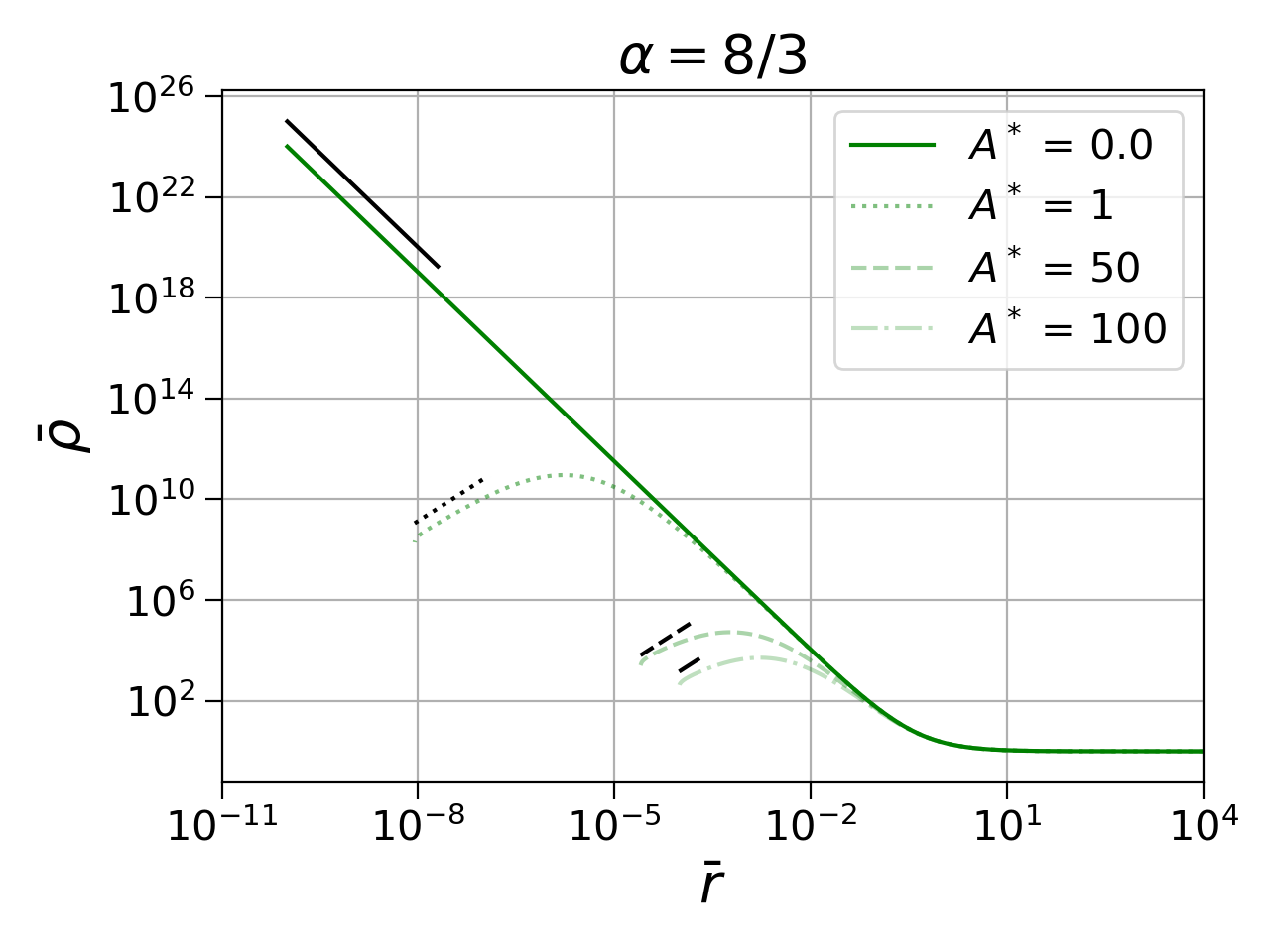}
\end{center}
\caption{Same as Fig.~\ref{Fig:subsonic}, but for density profiles.}
\label{Fig:subsonic_rho}
\end{figure*}

While we believe that, when it exists, supersonic accretion onto black holes is the most
likely astrophysically (see, e.g., Appendix G in \cite{ShaT83}, which
     shows that subsonic flow as gas approaches a black hole is ruled out
     in a general relativistic treatment of the adiabatic problem and that
     the flow will be driven supersonic), we also 
consider effects of heating on subsonic flows in
this Section.  As before, we will treat the cases $\alpha <2$, $\alpha = 2$, 
and $\alpha > 2$ separately. 

To construct subsonic solutions, we pick an accretion rate less than the corresponding
transonic accretion rate, $\dot{\bar M} < \dot{\bar M}_s$.  We also pick a large radius 
$\bar r_{\rm init} \gg \bar r_0$ and assume $\bar K \simeq 1$ there.  We express $\bar u$ in
terms of (\ref{M_dot_nd}), $\bar a$ in terms of (\ref{a_nd}) and insert these into the Bernoulli
equation (\ref{Bernoulli_nd}), yielding an equation for $\bar \rho$ at $\bar r_{\rm init}$.  With
these initial values, we then integrate (\ref{u_prime_nd}), (\ref{rho_prime_nd}) and 
(\ref{K_prime_nd}) inwards from $\bar r_{\rm init}$.

%===============================================================================
\subsection{Case 1: $\alpha < 2$}
\label{sec:subsonic_1}
%===============================================================================

We show an example for subsonic flow with $\alpha < 2$ in the left
panel of Fig.~\ref{Fig:subsonic}.  In this case, the fluid profiles
appear to approach the same power-law behavior for $\bar r \rightarrow
0$ as in the adiabatic case.  This behavior can be understood from the
following arguments.  Starting with the Bernoulli equation
(\ref{Bernoulli_nd}), we assume subsonic flow with $\bar u \ll \bar a$
as well as $\bar a \gg 1$ (i.e.~$a \gg a_\infty$).  The equation will
be dominated by the gravitational term at small $\bar r$ when $\alpha
< 2$, and the heating term can be neglected.  We therefore have
\begin{equation} \label{a_sub_1}
\bar a^2 \simeq \frac{\gamma - 1}{\bar r}, \hfill (\bar r \rightarrow 0,~\alpha < 2)
\end{equation}
just like in the adiabatic case (see Eq.~(14.3.28) in ST).  Inserting
this into (\ref{a_nd}) we obtain
\begin{equation} \label{rho_sub_1a}
\bar \rho = \left( \frac{\bar a^2}{\bar K} \right)^{1/(\gamma - 1)} 
\simeq \left( \frac{\gamma - 1}{\bar K \bar r} \right)^{1/(\gamma - 1)}, \hfill (\bar r \rightarrow 0,~\alpha < 2)
\end{equation}
(compare Eq.~(14.3.29) in ST).  In order to find an asymptotic scaling
for $\bar K$ we now insert (\ref{rho_sub_1a}) into (\ref{K_prime_nd})
to find
\begin{equation}
\bar K' \simeq - \gamma \frac{\bar A^* \bar r_{\rm ann}^\alpha \bar K}{\bar r^{\alpha - 1}}. \hfill (\bar r \rightarrow 0,~\alpha < 2)
\end{equation}
Integration yields
\begin{equation} \label{K_sub_1}
\bar K \propto \exp \left(- \gamma \bar A^* \bar r_{\rm ann}^\alpha \bar r^{2 - \alpha} / (2 - \alpha) \right), \hfill (\bar r \rightarrow 0,~\alpha < 2)
\end{equation} 
so that $\bar K$ approaches a (finite) constant as $\bar r \rightarrow
0$.  Inserting this result back into (\ref{rho_sub_1a}) we now have
\begin{equation} \label{rho_sub_1}
\bar \rho \propto \bar r^{-1/(\gamma - 1)}, \hfill (\bar r \rightarrow 0,~\alpha < 2)
\end{equation}
and, using the accretion rate (\ref{M_dot_nd}),
\begin{equation} \label{u_sub_1}
\bar u \propto \bar r^{-(2 \gamma - 3)/(\gamma - 1)},~~~~\hfill ~~(\bar r \rightarrow 0,~\alpha < 2)
\end{equation}
(see Eq.~(14.3.30) in ST).  For $\alpha < 2$ we therefore expect the
exact same power-law behavior for $\bar r \rightarrow 0$ as in the
adiabatic case.  For $\gamma = 5/3$, in particular, we recover the
free-fall behavior $\bar u \propto \bar r^{-1/2}$ and $\bar \rho \propto \bar r^{-3/2}$.  
Even in this case,
$\bar u$ and $\bar a$ increase with the same power law, meaning that a
solution with $\bar u < \bar a$ will remain subsonic.  We show
examples of this behavior in the left panels in
Figs.~\ref{Fig:subsonic} and \ref{Fig:subsonic_rho}, where the expected power laws are marked by
the black lines.

%===============================================================================
\subsection{Case 2: $\alpha = 2$}
\label{sec:subsonic_2}
%===============================================================================

We find very different asymptotic behavior in the special case $\alpha
= 2$.  In this case, the heating term scales with the same power as
the gravitational term in the Bernoulli equation (\ref{Bernoulli_nd}),
so that, considering the same limit as before, we now obtain
\begin{equation} \label{a_sub_2}
\bar a^2 \simeq (\gamma - 1) \frac{1 + \bar A^* \bar r_{\rm ann}^2}{\bar r}  \hfill (\bar r \rightarrow 0,~\alpha = 2)
\end{equation}
instead of (\ref{a_sub_1}).  From (\ref{a_nd}) we now have
\begin{equation} \label{rho_sub_2a}
\bar \rho % = \left( \frac{\bar a^2}{\bar K} \right)^{1/(\gamma - 1)} 
\approx \left( (\gamma - 1) \frac{1 + \bar A^* \bar r_{\rm ann}^2} {\bar K \bar r} \right)^{1/(\gamma - 1)}, \hfill (\bar r \rightarrow 0,~\alpha = 2)
\end{equation}
which we insert into (\ref{K_prime_nd}) 
\begin{equation}
\bar K' \simeq - \gamma \frac{\bar A^* \bar r_{\rm ann}^2}{1 + \bar A^* \bar r_{\rm ann}^2}  \frac{\bar K}{\bar r}. \hfill (\bar r \rightarrow 0,~\alpha = 2)
\end{equation}
Integration now yields
\begin{equation} \label{K_sub_2}
\bar K \propto \bar r^{-\gamma \delta}, \hfill (\bar r \rightarrow 0,~\alpha = 2)
\end{equation}
where we have abbreviated 
\begin{equation} \label{delta}
\delta \equiv \frac{\bar A^* \bar r_{\rm ann}^2}{1 + \bar A^* \bar r_{\rm ann}^2}.
\end{equation}
Inserting (\ref{K_sub_2}) into (\ref{rho_sub_2a}) now yields
\begin{equation} \label{rho_sub_2}
\bar \rho \propto \bar r^{-(1 - \gamma \delta)/(\gamma - 1)}, \hfill (\bar r \rightarrow 0,~\alpha = 2) 
\end{equation}
and, using (\ref{M_dot_nd}) again,
\begin{equation} \label{u_sub_2}
\bar u \propto \bar r^{-(2\gamma - 3 + \gamma \delta)/(\gamma - 1)}.  \hfill (\bar r \rightarrow 0,~\alpha = 2) 
\end{equation}
Interestingly, the power-law scaling now depends on the heating rate
$\bar A^*$ through $\delta$.  We show examples for this behavior in
the middle panels of Figs.~\ref{Fig:subsonic} and \ref{Fig:subsonic_rho}, where we again find
excellent agreement between our numerical result and the power-law
behavior expected from the above arguments.  Note that we have $\delta
\rightarrow 0$ in the adiabatic limit, in which case our results above
reduce to those of Case 1 in Section \ref{sec:subsonic_1}, as expected.
For sufficiently small heating rate, and hence sufficiently small
$\delta$, the fluid velocity $\bar u$ still grows more slowly than the
sound speed $\bar a$ as $\bar r \rightarrow 0$, so that a subsonic
solution will remain subsonic.  For
\begin{equation}
\delta > \delta_{\rm crit} = \frac{5 - 3 \gamma}{2 \gamma},
\end{equation}
corresponding to a heating rate
\begin{equation}
\bar A^* > \bar A^*_{\rm crit} =  \frac{5 - 3 \gamma}{5 (\gamma - 1)} \,\bar r_{\rm ann}^{-2},
\end{equation}
however, $\bar u$ increases more rapidly than $\bar a$ as $\bar r
\rightarrow 0$, suggesting that this solution will {\em not} remain
subsonic for arbitrarily small $\bar r$.  This contradicts our
assumption $\bar u \ll \bar a$, of course, so that our approximations
will no longer remain accurate.  We also caution that, for DM heating,
the exponent $\alpha$ would probably drop to a smaller value at $\bar
r \sim \bar r_{\rm ann}$ (see Section \ref{sec:heating}), which we
ignored in our treatment here.  The appearance of a critical heating
rate is reminiscent of that for transonic flow with $\alpha = 2$ in
Section \ref{sec:find_r_s_equal_2}.

%===============================================================================
\subsection{Case 3: $\alpha > 2$}
\label{sec:subsonic_3}
%===============================================================================

Finally we consider the case $\alpha > 2$.  Making the same
assumptions of $\bar u \ll \bar a$ and $\bar a \gg 1$ as before in the
Bernoulli equation (\ref{Bernoulli_nd}), we now see that the heating
term dominates at small $\bar r$, so that we may approximate
\begin{equation} \label{a_sub_3a}
\frac{\bar a^2}{\gamma - 1} \approx \frac{\bar A^*}{\alpha - 1} \frac{\bar r_{\rm ann}^\alpha}{\bar r^{\alpha - 1}}
  \hfill (\bar r \rightarrow 0,~\alpha > 2) 
\end{equation}
(instead of (\ref{a_sub_1}) and (\ref{a_sub_2})).  From (\ref{a_nd})
we then have
\begin{equation} \label{rho_sub_3a}
\bar \rho  % = \left( \frac{\bar a^2}{\bar K} \right)^{1/(\gamma - 1)} 
\approx \left( \frac{\gamma - 1}{\alpha - 1} \frac{\bar A^*}{\bar K} \frac{\bar r_{\rm ann}^{\alpha}}{\bar r^{\alpha - 1}} 
\right)^{1/(\gamma - 1)}.   \hfill (\bar r \rightarrow 0,~\alpha > 2) 
\end{equation}
Inserting (\ref{rho_sub_3a}) into (\ref{K_prime_nd}) yields
\begin{equation}
\bar K' \approx - \gamma (\alpha - 1) \frac{\bar K}{\bar r},   \hfill (\bar r \rightarrow 0,~\alpha > 2) 
\end{equation}
which we can integrate to obtain
\begin{equation} \label{K_sub_3}
\bar K \propto \bar r^{- \gamma (\alpha - 1)}.   \hfill (\bar r \rightarrow 0,~\alpha > 2) 
\end{equation}
As before, we now insert (\ref{K_sub_3}) back into (\ref{rho_sub_3a}) to find
\begin{equation}
\bar \rho \propto \bar r^{\alpha - 1}, \hfill (\bar r \rightarrow 0,~\alpha > 2) 
\end{equation}
and combine this with the accretion rate (\ref{M_dot_nd}) to find 
\begin{equation} \label{u_sub_3}
\bar u \propto \bar r^{-(\alpha + 1)}.  \hfill (\bar r \rightarrow 0,~\alpha > 2)
\end{equation}
Note that the power law exponents for the fluid variables $\bar a$,
$\bar \rho$ and $\bar u$ are independent of both $\gamma$ and the
heating rate in this case, and instead depend on $\alpha$ only.  Also
note that, for all $\alpha > 2$, $\bar u$ increases more rapidly than
\begin{equation} 
\bar a \propto \bar r^{-(\alpha - 1)/2} \hfill
(\bar r \rightarrow 0,~\alpha > 2)
\end{equation}
with decreasing $\bar r$.  While our estimates assume that $\bar u \ll
\bar a$, they again suggest that this assumption will break down at
some sufficiently small $\bar r$, once the heating term dominates.  In
fact, these results suggest that ``subsonic" solutions may not remain
subsonic to arbitrarily small radii, instead they may encounter a
sonic point at some some radius $\bar r$, where $\bar u = \bar a$.
This is exactly what our numerical explorations of this regime
suggest.  We show examples in the right panels of
Figs.~\ref{Fig:subsonic} and \ref{Fig:subsonic_rho}, where we have also included the expected
power-law behavior.  As one might expect, for larger values of $\bar
A^*$ the flow will deviate from the adiabatic flow, and be dominated
by the heating term, starting at larger values of $\bar r$.  For small
heating we find very good agreement between the numerical results and
the expected power law, while for larger heating the assumption $\bar
u \ll \bar a$ appears to be violated before $\bar u$ can approach the
heating-dominated power law.

For generic accretion rate, the sonic point found in this process will
not satisfy the conditions laid out in Section
\ref{sec:transonic_strategy}; in particular the numerators and
denominators on the right-hand sides of Eqs.~(\ref{u_prime_nd}) and
(\ref{rho_prime_nd}) will not have simultaneous roots, so that these
solutions will not describe smooth fluid flow.

Combining this finding with that of Section
\ref{sec:find_r_s_greater_2} we conclude that, for $\alpha > 2$, we
can find neither supersonic nor subsonic solutions that describe
smooth, steady-state spherical accretion for all radii.  We will
comment on this result, as well as its limitations, in more detail in
Section \ref{sec:summary}.  In particular, we remind the reader that
we have assumed a constant $\gamma_{\rm sp}$ for all $r$ in the DM
density distribution (\ref{spike}), whereas we would expect
$\gamma_{\rm sp}$ to switch to $\gamma_{\rm ann}$ at $r_{\rm ann} \sim
r_{\rm a} / 20$.  Clearly, relaxing this assumption will affect the
findings for very small $r$ in this section.

%===============================================================================
\section{Applications to Sgr A$^*$}
\label{sec:Sgr_A}
%===============================================================================

\begin{figure}
\begin{center}
\includegraphics[width=0.45 \textwidth]{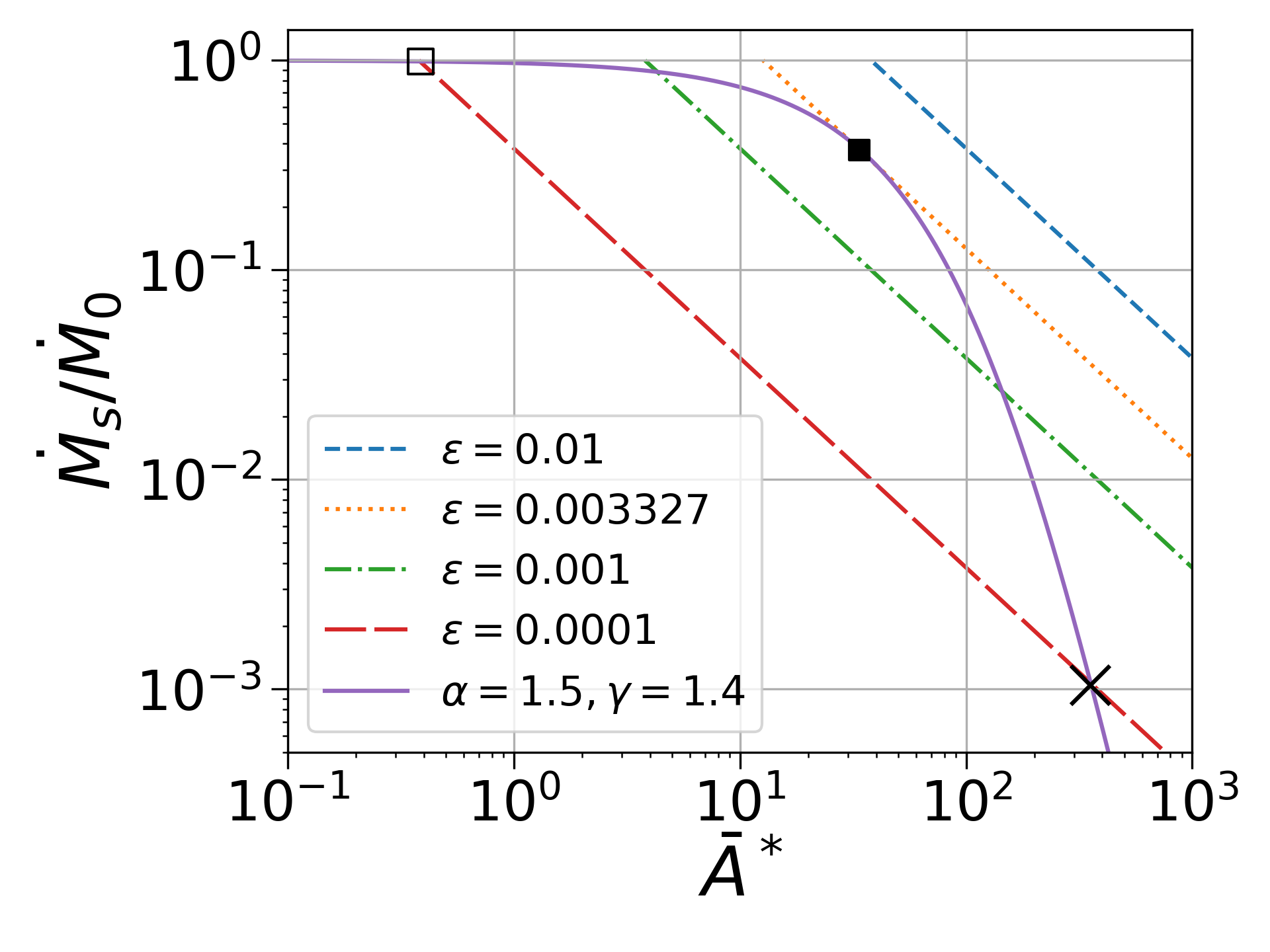}
\end{center}
\caption{The accretion rate $\dot M/ \dot M_0$ as a
  function of the heating parameter $\bar A^*$, for $\alpha = 1.5$ and
  $\gamma = 1.4$. The solid (purple) line represents numerical results for
  {\em transonic} solutions (see Section \ref{sec:transonic_results}), while
  the straight lines represent the hyperbolae
  (\ref{M_dot_hyperbola}) for efficiencies $\epsilon =$ 0.01, 0.001, 0.003327, and
  0.0001.  Intersections of the solid line with the hyperbolae
  represent viable solutions for the DM parameters assumed in
  (\ref{C}), and identify the associated accretion rates $\dot M /
  \dot M_0$. The solutions identified by the solid square and the cross, for example,
  represent spherically symmetric, steady-state accretion for which
  the heating by DM annihilation has reduced the accretion rate by a
  factor of about $0.37$ and about $1 \times 10^{-3}$ respectively.   For the solution marked by the 
  open square, which corresponds to the same heating efficiency as that marked by the cross,
  the accretion rate is reduced by a factor of $0.988$ only (see text for a discussion).
  We show flow 
  profiles for the solutions represented by the cross and the two squares in Fig.~\ref{Fig:profile}.}
\label{Fig:M_dot2}
\end{figure}

\begin{figure}
\begin{center}
\includegraphics[width=0.45 \textwidth]{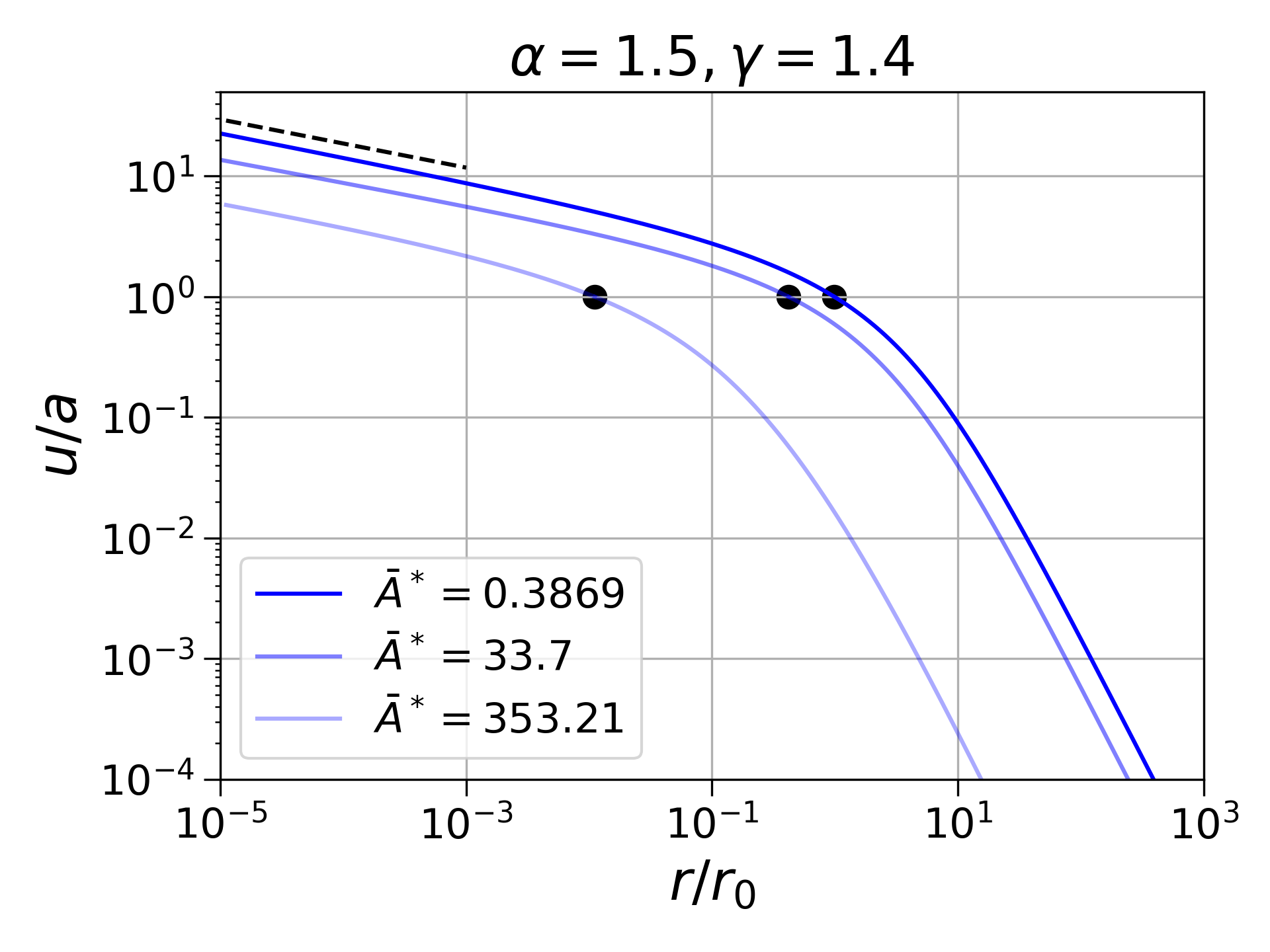}
\includegraphics[width=0.45 \textwidth]{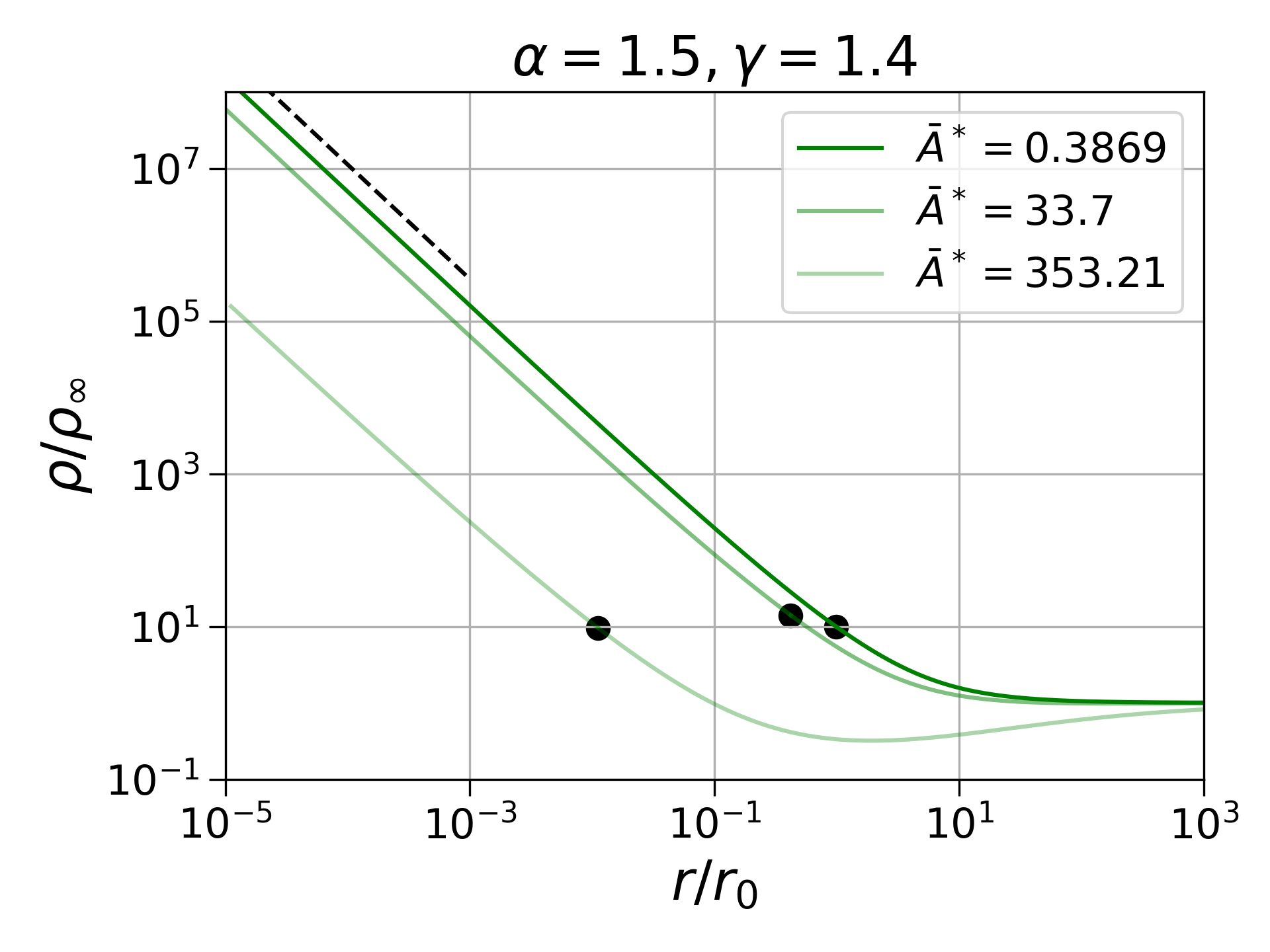}
\end{center}
\caption{Fluid flow profiles for the heated, transonic accretion
  solutions for $\gamma = 1.4$ and $\alpha = 1.5$ and marked by the 
  cross, and the solid and open squares in Fig.~\ref{Fig:M_dot2}. For the 
  solution denoted by the cross in Fig.~\ref{Fig:M_dot2} the accretion rate 
  is reduced by a factor of about $1 \times 10^{-3}$ below the adiabatic 
  Bondi accretion rate. For this solution, displayed as the lightest solid line,
  $\bar{A}* = 353.21$. 
  For the solution denoted by the solid (open) square in Fig.~\ref{Fig:M_dot2}, 
  for which $\bar{A}^* = 33.7$ ($\bar A^* = 0.3869$), the accretion rate is 
  reduced by a factor of about $0.37$ ($0.988$) below the corresponding 
  adiabatic Bondi accretion rate. For Sgr A$^*$, 
  $\rho_\infty \approx 3 \times 10^{-23}$ g cm$^{-3}$, and $r_s \approx 0.06$ pc. 
  Included in the fluid profiles are the expected power laws for the transonic fluid 
  velocity, $u$ $\approx r^{-1/2}$, and fluid density, $\rho$ $\approx r^{-3/2}$ 
  (see Eqs.~(14.3.24) and (14.3.26) in ST). }
\label{Fig:profile}
\end{figure}

In this Section we explore whether, for reasonable choices of DM
parameters, heating by DM annihilation could explain the low accretion
rates observed for Sgr A$^*$ in the GC, with $\dot M / \dot M_0 \sim
10^{-3}$.  Our estimates in Section \ref{sec:heating} suggest that DM
annihilation may have an order unity effect, and we will now
re-examine these effects in the context of transonic solutions for
simple Bondi accretion.

In order to evaluate our results quantitatively for DM parameters
considered realistic for the environment of Sgr A$^*$ in the GC, we
first need to express the heating parameter $\bar A^*$ in terms of the
DM parameters.  This is complicated by the fact that $A^*$, and hence
the nondimensional version $\bar A^*$, depends on the accretion rate
$\dot M$ (see eq.~(\ref{A_star})), which, in turn, is a result of a
calculation for a given value of $\bar A^*$.  In order to disentangle
these dependencies we use (\ref{A_star_nd}) and (\ref{A_star}) to
write
\begin{equation} \label{A_star_2}
\bar A^* = \frac{r_a}{a_\infty^2} A^* = \frac{4 \pi \Gamma_0 r_{\rm ann}^2 r_a}{a_\infty^2 \dot M} =
\frac{4 \pi \Gamma_0 r_{\rm ann}^2 r_a}{a_\infty^2 \dot M_0}  \left( \frac{\dot M_0}{\dot M} \right).
\end{equation}
We now define the dimensionless quantity
\begin{equation} \label{C}
{\mathcal C} = \frac{4 \pi \Gamma_0 r_{\rm ann}^2 r_a}{a_\infty^2 \dot M_0} 
\end{equation}
and evaluate, for the canonical parameters of Section
\ref{sec:heating},
${\mathcal C} \sim \epsilon \times 3.8 \times 10^3$.  We can then
solve (\ref{A_star_2}) for $\dot M / \dot M_0$ to find
\begin{equation} \label{M_dot_hyperbola}
\frac{\dot M}{\dot M_0} = \frac{\mathcal C}{\bar A^*}.
\end{equation}
For a given value of ${\mathcal C}$, the computed accretion rate $\dot
M / \dot M_0$ has to agree with that found from
(\ref{M_dot_hyperbola}).  In practice, we look for intersections of the
hyperbolae (\ref{M_dot_hyperbola}) with our computed accretion rates, as
shown in Fig.~\ref{Fig:M_dot2}.  Given our findings in Section
\ref{sec:transonic_results} we focus on $1 < \alpha < 2$ and $1 <
\gamma < 5/3$ in Fig.~\ref{Fig:M_dot2}.

As an aside, we note that we can also express ${\mathcal C}$ as 
\begin{equation}
{\mathcal C} = \frac{4 \pi \Gamma_0 r_a^3}{a_\infty^2 \dot M_0} \left( \frac{r_{\rm ann}}{r_a} \right)^2
= \frac{4 \pi \Gamma_0 r_a^3}{a_\infty^2 \dot M_0} \left( \frac{r_{\rm ann}}{r_a} \right)^{2 \gamma_{\rm sp}} \left( \frac{r_{\rm ann}}{r_a} \right)^{-\alpha}
\end{equation}
and, up to a difference between $a(r_a)$ and $a_\infty$, recognize the
first two terms on the right-hand side as the ratio between the
heating rate and the rate of thermal energy flow (see
eq.~(\ref{ratio})) evaluated at $r = r_a$, so that
\begin{equation}
{\mathcal C} \sim {\cal R}(r_a) \bar r_{\rm ann}^{-\alpha}.
\end{equation}
Accordingly, we may also write (\ref{M_dot_hyperbola}) as 
\begin{equation}
\frac{\dot M}{\dot M_0} \sim \frac{{\cal R}(r_a) }{\bar A^* \bar r_{\rm ann}^\alpha}.
\end{equation}

Returning to Fig.~\ref{Fig:M_dot2}, we note that there do indeed exist
viable transonic solutions for which DM heating reduces spherical
Bondi accretion to small values.  A specific example for which the 
accretion rate is reduced by three orders of magnitude below the corresponding
Bondi value is marked by the cross in
Fig.~\ref{Fig:M_dot2}.  In Fig.~\ref{Fig:profile} we explore this
solution in more detail, and show the fluid flow profiles as a
function of radius.

We caution, however, that our solutions represent equilibrium solutions that may or may not be stable. In Fig.~\ref{Fig:M_dot2} we see that, if the hyperbolae (\ref{M_dot_hyperbola}) intersect the computed accretion rate for a given efficiency $\epsilon$, then there are two intersections corresponding to two viable equilibrium solutions.  For $\epsilon = 10^{-4}$, for example, we have marked these two intersections with an open square and a cross in Fig.~\ref{Fig:M_dot2}.  While this figure shows results for $\alpha = 1.5$ and $\gamma = 1.4$ only, we have found similar behavior for all parameters that we have considered.   It is possible that these two solutions represent members of a stable and an unstable branch of solutions, separated by the point at which the computed accretion rate curve is tangent to the hyperbolae (\ref{M_dot_hyperbola}).  In Fig.~\ref{Fig:M_dot2} we marked this point with the solid square.  The two branches behave differently as we reduce the heating efficiency.  For the upper branch (on which the open square is located) the accretion rate approaches the Bondi rate when the efficiency is lowered (and hence the heating rate decreases), while for the lower branch (on which the cross is located) the accretion rate decreases.  This suggests that the upper branch may represent stable equilibria, while the lower branch may represent unstable equilibria.   Establishing the stability properties of these branches would require either a perturbative treatment or dynamical numerical simulations, both of which are beyond the scope of this paper.   If indeed only the upper branch of solutions in Fig.~\ref{Fig:M_dot2} were stable, then this stable branch would end with the marginally stable, critical solution marked by the solid square.  We have included fluid flow profiles for this (possibly) critical solution in Fig.~\ref{Fig:profile}.

We also note that even equilibrium solutions, irrespective of their stability, exist only for a limited
range of parameters, and not necessarily for those parameters that are
favored on astrophysical grounds.  In particular, no such solutions
exist for $\gamma = 5/3$ (even though the lack of solutions for
$\gamma = 5/3$ might be an artifact of our Newtonian treatment of the
problem, {\it cf.} Appendix G in ST), nor can we find regular solutions
for $\alpha > 2$ ($\gamma_{\rm sp} > 2$).  Our results nevertheless confirm our expectation,
based on the estimates in Section \ref{sec:heating}, that heating by
DM annihilation may play an important role in other more detailed
accretion flows.

%===============================================================================
\section{Summary and Discussion}
\label{sec:summary}
%===============================================================================

We examined effects of heating by DM annihilation on spherical
accretion onto black holes.  Adopting
plausible values for DM densities, as well as DM masses and annihilation
cross-sections within the WIMP model, we estimate that such heating
may have an order unity effect on accretion onto Sgr A$^*$ in the GC.
If indeed present, such heating may therefore play an important role
for these accretion processes, and may, in fact, help explain the low
accretion rate observed for Sgr A$^*$.

Motivated by this observation we studied the effects of heating
on the simplest possible
accretion model, namely spherically symmetric, steady-state Bondi flow
of a gas with adiabatic index $\gamma$.  For many choices of the DM density 
spike power-law parameter $\alpha$ and the parameter $\gamma$, 
including those that are probably
favored on astrophysical grounds, we do not find smooth transonic
solutions.  For other parameters, however, we do find such solutions.
In particular, we present in Section \ref{sec:Sgr_A} as an ``existence
proof" some viable solutions with low accretion rates that may model
accretion flow onto Sgr A$^*$.

Evidently, our discussion is affected by many assumptions, and
therefore comes with many caveats.  For starters, we have assumed
certain canonical values for DM and Galactic parameters.  Some of
these parameters are based on observational data, but others are very
uncertain -- including the DM particle mass and cross-sections and the
efficiency $\epsilon$ with which energy generated by particle
annihilation ends up heating the accreting gas.

Moreover, our treatment of accretion within the Bondi model assumes
smooth, spherically symmetric and steady-state flow onto the black
hole, which presumably is also not realistic.  While we believe that it
is useful to explore the effects of heating by DM annihilation within
this simple model, its predictability for the GC is, of course, 
limited.  Conservation of angular
momentum may change the flow from near-radial infall to disk-like
accretion at small radii, so that the singular behavior that we find
for radial flow at small radii may not be realized in more realistic
situations.  On the other hand, our results suggest that, for many
values of $\alpha$ and $\gamma$, strictly spherical, smooth
steady-state accretion in the presence of heating (described by a
single power law) does not exist.  Even in these cases, accretion might
still be possible, but it would have to violate at least one of the
assumptions made: it could be episodic rather than steady-state, it
could feature shocks (especially at the inner sonic radius) rather
than being smooth, or it may break spherical
symmetry.  In any case, our results already suggest that the effects
of heating by DM annihilation should be considered in future, more
detailed hydrodynamic simulations of gas flow onto Sgr A$^*$.

\section*{Acknowledgments}

We thank C.~Gammie, B.~Fields and J.~Shelton for helpful discussions.
ERB acknowledges support through an undergraduate research fellowship
at Bowdoin College.  This work was supported in part by NSF grant
PHYS-1707526 to Bowdoin College,  NSF
grant  PHY-1662211 and NASA grant 
80NSSC17K0070 to the University of Illinois at Urbana-Champaign, as
well as through sabbatical support from the Simons Foundation (Grant
No.~561147 to TWB).

\bibliographystyle{mnras}
% \bibliography{references}

\end{document}